\documentclass{article}

\usepackage[preprint]{neurips_2026}

\usepackage[utf8]{inputenc} % allow utf-8 input
\usepackage[T1]{fontenc}    % use 8-bit T1 fonts
\usepackage{amsmath}
\usepackage{hyperref}       % hyperlinks
\usepackage[capitalize,noabbrev]{cleveref}
\usepackage{xspace}

\newcommand{\Trak}{\textsc{Trak}\xspace}
\newcommand{\FactGraSS}{\textsc{FactGraSS}\xspace}
\newcommand{\LoGra}{\textsc{LoGra}\xspace}
\newcommand{\GradCos}{\textsc{Grad-Cos}\xspace}

\newcommand{\CLAP}{CLAP\xspace}
\newcommand{\CLEWS}{CLEWS\xspace}
\newcommand{\MERT}{MERT\xspace}

\newcommand{\ARIA}{\textsc{Aria}\xspace}

\Crefname{equation}{Eq.\!}{Eqs.\!}
\Crefname{table}{Table}{Tables}
\Crefname{figure}{Figure}{Figures}

\usepackage{caption}
\usepackage{url}            % simple URL typesetting
\usepackage{booktabs}       % professional-quality tables
\usepackage{multirow}
\usepackage{amsfonts}       % blackboard math symbols
\usepackage{nicefrac}       % compact symbols for 1/2, etc.
\usepackage{microtype}      % microtypography
\usepackage{xcolor}         % colors
\usepackage{enumitem}       % list customization
\usepackage{float}          % [H] placement specifier
\usepackage{wrapfig}        % wraptable/wrapfigure
\usepackage{placeins}       % \FloatBarrier
\usepackage{graphicx}
\usepackage{pgfplots}
\pgfplotsset{compat=1.18}
\usepackage{subcaption}
\usepackage{tikz}
\usepgfplotslibrary{groupplots}

\title{\ARIA: A Diagnostic Framework for Music Training Data Attribution}

\author{%
  Changheon Han$^1$ \quad Ashkan Panahi$^1$ \quad Kıvanç Tatar$^1$\thanks{Corresponding author}
   \\ 
  $^1$Department of Computer Science and Engineering,\\
  Chalmers University of Technology {and} University of Gothenburg\\
  Gothenburg, Sweden\\  
  \texttt{\{changheon.han, ashkan.panahi, tatar\}@chalmers.se}
}

\begin{document}

\maketitle

\begin{abstract}
Training data attribution (TDA) for music generation must answer two questions that copyright analysis requires, namely which training songs influence a generated output and along which musical aspects the influence operates. Existing methods reduce influence to a single scalar, without revealing which musical aspects are dominant in that influence. We propose \ARIA, a framework that decomposes attribution along musical aspects (five for symbolic music, three for audio) and pairs the decomposition with reliability diagnostics computed from the segment-level score matrix. It measures within-group similarity among the top-$K$ attributed tracks against random reference groups drawn from the training pool, and diagnoses the score matrix through its singular value decomposition and column statistics. On a symbolic-music model where attribution ground truth is available through counterfactual retraining, the reliability diagnostics rank four attribution methods identically to that ground truth. On an audio music generation model, \ARIA reveals attribution behaviors that vary substantially across TDA methods, flags score matrices whose retrieved tracks are nearly identical across queries rather than reflecting per-query attribution, and characterizes embedding-similarity retrieval baselines by the musical aspect each encoder surfaces. Together, \ARIA produces per-aspect attribution evidence aligned with the musical aspects considered under the idea-expression distinction in copyright analysis.
\end{abstract}

% ============================================================
\section{Introduction}
% ============================================================

In January 2026, folk musician Murphy Campbell discovered AI-generated covers of her own performances on Spotify, derived from her YouTube videos and uploaded under her name without her knowledge or consent. Despite weeks of repeated reports, the same songs were re-uploaded multiple times under different listings~\citep{theverge2026campbell}. Such cases show how easily an artist's work can be exploited through generative models. In broader contexts, major record labels have filed lawsuits against commercial music generation systems over unauthorized use of copyrighted audio recordings~\citep{umg2024suno,umg2024udio}. Generative models can memorize and reproduce training data~\citep{carlini2023extracting,carlini2021extracting,somepalli2023diffusion}, yet the technical problem of tracing which training songs shape a model's synthetic outputs, and along which musical aspects, remains largely unexplored.

Tracing training data influence is notably different in music than in text or image domains, where memorization is detected through verbatim matching or pixel-level retrieval~\citep{carlini2023extracting,somepalli2023diffusion,carlini2023quantifying}. Music similarity spans multiple entangled musical aspects such as melody, harmony, rhythm, and timbre \cite{herremans2017functional}, and no single aspect characterizes the influence~\citep{lee2020disentangled,lee2020metric}. Legal analysis reflects the same multi-aspect structure. Under the idea-expression distinction, music infringement is evaluated across multiple aspects and protects specific expressive elements rather than underlying stylistic ideas~\citep{mullensiefen2009court,dornis2025copyright}. Meaningful attribution must therefore identify which training songs are influential and along which musical aspects that the influence operates.

Existing training data attribution (TDA) methods~\citep{koh2017understanding,pruthi2020estimating,ilyas2022datamodels,park2023trak} reduce each training example's influence to a single scalar, leaving the per-aspect question unanswered. Music-specific attempts inherit this gap. Compensation designs and unlearning-based evaluations~\citep{deng2023computational,choi2025large,kim2025noencore} produce one number per track. Embedding retrieval~\citep{barnett2024exploring} ranks tracks by encoder cosine similarity and inherits whichever aspects the encoder happens to encode. Attribution-by-design~\citep{morreale2025attribution} requires provenance be built into the system at training time, which existing generative music models do not support.

We introduce \ARIA (\textbf{A}ttribution \textbf{R}esult \textbf{I}nterpretation and \textbf{A}nalysis), a framework that reads attribution outputs along the musical aspects copyright analysis requires. \ARIA has two components. The first component measures within-group similarity among the top-$K$ attributed tracks along musical aspect channels and standardizes each channels against random reference groups drawn from the training pool, surfacing whether a method's signal is concentrated on certain aspects over others. The channel set is matched to the evaluation domain. For symbolic domain, exact MIDI permits a finer decomposition into melody, harmony, rhythm, dynamics, and texture, the common musical elements considered in copyright analysis~\citep{mullensiefen2009court,dornis2025copyright}. We use the dattri benchmark~\citep{deng2024dattri} to obtain a pre-trained MusicTransformer~\citep{huang2019music} on MAESTRO~\citep{hawthorne2018maestro} with a published linear datamodeling score (LDS)~\citep{ilyas2022datamodels} as attribution ground truth. For audio domain, we use rhythm, harmony, and timbre, instantiate on a MusicLM-style~\citep{agostinelli2023musiclm,borsos2023audiolm} three-stage hierarchical musical audio generation model trained on FMA Large~\citep{defferrard2017fma}, where LDS is computationally infeasible at scale. The second component diagnoses the segment-level score matrix through three structural quantities derived from its singular value decomposition~\citep{mardia1979multivariate} and column statistics, detecting when an attribution method assigns nearly the same ranking to every query so that per-aspect results characterize a fixed retrieved group rather than per-query attribution. 

Our contributions are in four threads: 

\begin{enumerate}[leftmargin=1.5em,itemsep=2pt,topsep=2pt]
\item \ARIA surfaces per-aspect attribution evidence along the common musical aspects for copyright analysis already uses, supporting compensation and infringement analysis that scalar scores cannot.
\item We identify two structural confounds in per-aspect music attribution. Aspect imbalance concentrates attribution signal on certain musical aspects over others, and query-independent score collapse causes attribution methods to rank training segments nearly identically across queries.
\item On a symbolic music generation model with attribution ground truth, \ARIA's reliability diagnostics rank four attribution methods identically to that ground truth.
\item On a musical audio generation model where counterfactual retraining is infeasible, \ARIA reveals attribution profiles that vary substantially across methods, and flags several score matrices whose per-aspect results reflect query-independent collapse rather than per-query attribution.
\end{enumerate}

% ============================================================
\section{Related Work}
% ============================================================

\subsection{Musical Similarity in Copyright Analysis}
\label{sec:related-copyright}
Music infringement analysis decomposes similarity into distinct aspects, with courts and musicologists routinely separating melody, harmony, and rhythm under the idea-expression distinction~\citep{livingston2013copyright,nicolas2023harmonizing,mullensiefen2009court,dornis2025copyright}. The same need extends to compensation mechanisms in generative music, where transparency over which works contribute is argued as a precondition for fair allocation~\citep{morreale2025attribution}. Music information retrieval treats similarity in the same way, with rhythm, harmony, and timbre developed as separate analytical axes through dedicated signal descriptors~\citep{muller2015fundamentals,mcfee2015librosa,tatar2021latent}. A complementary line pursues learned representations that disentangle these axes, through metric learning across semantic axes such as genre and instrumentation~\citep{lee2020disentangled,lee2020metric}, variational autoencoders for pitch and timbre on instrument-level audio~\citep{luo2019learning,esling2021flowsynth}, and self-supervised separation of harmonic and rhythmic features on full-mix audio~\citep{wu2023self}. Each covers only part of the space, so \ARIA adopts signal-level descriptors along the three aspects they reliably cover (rhythmic, harmonic, timbral) for audio domain leaving melody aside since polyphonic melody extraction from full-mix audio remains unresolved~\citep{salamon2014melody,bittner2017deep}.

\subsection{Training Data Attribution}

TDA is commonly evaluated against the LDS~\citep{ilyas2022datamodels,park2023trak}, which correlates attribution scores, summed over a random training subset, with the behavior of a model retrained on that subset. 
LDS requires retraining many models on different corpus subsets and is infeasible at the scale of modern generative models. 
Evaluation at this scale instead exploits modality-specific proxies, such as verbatim memorization~\citep{carlini2021extracting,carlini2023quantifying} and fact tracing~\citep{akyurek-etal-2022-towards} in text, or controlled customization, unlearning, and concept-level attribution in image~\citep{wang2023evaluating,wang2024unlearning,park2025concepttrak}. 
Music supports neither route, since musical similarity is multi-aspect and resists both verbatim matching and single-concept customization. 

Music-specific attribution approaches each rely on a distinct reference to evaluate method quality.
\citep{deng2023computational} confine evaluation to MIDI-rendered audio
where symbolic benchmark metrics apply, sidestepping audio attribution
rather than resolving it.
\citep{barnett2024exploring} score tracks by the cosine similarity of audio encoder embeddings,
supplemented with listening tests, inheriting which musical aspects the encoder encodes.
\citep{choi2025large} and \citep{kim2025noencore} simulate track removal
by updating model parameters per query, making each evaluation expensive.
\citep{morreale2025attribution} propose to build attribution provenance into the
model design itself, which existing generative models do not support.
None of these methods answer which musical aspects the influence operates on.

\ARIA diagnoses attribution quality directly from the score matrix,
without retraining cost, ground-truth labels, or external perspectives
that risk introducing biases. It further reveals which musical aspects a scoring method actually
captures, a question that single scalar score cannot expose.

% ============================================================
\section{\ARIA: A Diagnostic Framework for Music Attribution}
\label{sec:method}
% ============================================================

\ARIA evaluates an attribution score matrix from two complementary angles
without requiring ground-truth attribution labels.
We diagnose the reliability of the score matrix through three structural
quantities of its spectrum and column statistics (\Cref{sec:reliability}),
and characterize the musical homogeneity of the attributed group along
modality-specific evidence channels (\Cref{sec:homogeneity}).
We first instantiate \ARIA on a symbolic music generation model where LDS attribution
ground truth and exact musical similarity are both available, and then on
a musical audio generation model where neither is tractable at scale.

\subsection{Problem Setting}
\label{subsec:problem-setting}

Let $D = \{x_1, \ldots, x_N\}$ be a training corpus of music tracks and let
$f$ be a generative music model trained on $D$.
Each track $x_i$ is split into fixed-length segments
$\{s_{i,1}, \ldots, s_{i,n_i}\}$, where the segment unit is
modality-specific, such as a fixed-duration audio window or a fixed-length
symbolic token block.
Let $Q = \{q_1, \ldots, q_T\}$ be a set of queries derived from generated
outputs of $f$.
A \emph{scoring method} $\tau$ assigns a real-valued score
$\tau(q, s) \in \mathbb{R}$ to every training segment $s$ given a query
$q$, expressing how strongly $s$ is implicated in the production of $q$.
Two families of scoring methods are considered.
\textbf{Attribution methods} compute $\tau$ from training-time loss
gradients at each segment.
Alternatively, \textbf{embedding-based retrieval baselines} compute $\tau$ as the cosine
similarity between fixed audio embeddings of the generated output and the
training track. Concrete formulations of every method used in our
experiments appear in Appendix~\ref{app:attribution_formulations}.

The raw output of $\tau$ is the \emph{segment-level} score matrix
$S^{\mathrm{seg}} \in \mathbb{R}^{M \times T}$ with entries
$S^{\mathrm{seg}}_{(i,k),\, j} = \tau(q_j, s_{i,k})$, where
$M = \sum_{i=1}^{N} n_i$ is the total number of training segments.

For the homogeneity analysis we additionally form a
\emph{track-level} matrix $S^{\mathrm{track}} \in \mathbb{R}^{N \times T}$.
Because each segment $s_{i,k}$ must first be aggregated into a single
track-level score, and because raw per-query score ranges differ across
methods and queries such that a query with high-variance scores would
otherwise dominate the averaged result, each column of $S^{\mathrm{seg}}$
is first rescaled to a common scale through \emph{per-query normalization}.

When a method produces columns that are roughly symmetric and
free of extreme outliers, \textbf{Z-score per test query} is used,
standardizing each column to zero mean and unit variance.
When columns are heavy-tailed, Z-scoring lets a small number of
outlier segments dominate the per-track average. In this case,
\textbf{Rank per test query} is instead used, replacing each column
by its rank order, which is outlier-resistant by construction.
The choice of normalization for each scoring method used in our
experiments is specified in Appendix~\ref{app:query-norm}.

The normalized segment scores are then averaged within each track,
\begin{equation}
\label{eq:track-agg}
S^{\mathrm{track}}_{ij} \;=\; \frac{1}{n_i} \sum_{k=1}^{n_i}
\tilde{S}^{\mathrm{seg}}_{(i,k),\, j},
\qquad i \in \{1, \ldots, N\},\ j \in \{1, \ldots, T\}.
\end{equation}

For the symbolic domain experiment, training instances are indexed at the
segment level. Hence, we have $S^{\mathrm{track}} = S^{\mathrm{seg}}$.
For embedding baselines, the encoder output is time-averaged into one
vector per track before scoring, so that the score matrix is
naturally track-indexed in $\mathbb{R}^{N \times T}$.

% ============================================================
\subsection{Score-Matrix Reliability Diagnostics}
\label{sec:reliability}
% ============================================================

Reliable per-query attribution requires that the scoring method assign meaningfully different rankings to different queries. We diagnose this with three structural quantities of the segment-level score matrix $S^{\mathrm{seg}}$, each isolating a distinct way the matrix can deviate from query-dependent ranking.

\paragraph{Mean absolute inter-query correlation.}
The column $S^{\mathrm{seg}}_{\cdot j}$ records the per-segment score
profile that $\tau$ assigns under query $q_j$, so two columns are
highly correlated when their queries induce the same score profile
across training segments up to an affine transformation.
Let $C \in \mathbb{R}^{T \times T}$ be the Pearson correlation matrix~\cite{lee1988thirteen}
between the columns of $S^{\mathrm{seg}}$. The mean absolute inter-query
correlation is
\begin{equation}
\label{eq:kappa}
\kappa = \frac{1}{T(T-1)} \sum_{j \neq j'} |C_{jj'}| \in [0, 1].
\end{equation}
$\kappa$ close to $1$ means almost every pair of queries produces the
same score profile up to an affine transformation, so query input is
largely ignored.

\paragraph{Singular value energy ratios.}
While $\kappa$ captures pairwise column similarity directly, a
complementary question is whether the score matrix is dominated by a
single global axis, a failure mode with a distinct spectral signature
that motivates the following decomposition.
The singular value decomposition (SVD) of $S^{\mathrm{seg}}$ writes
$S^{\mathrm{seg}} = \sum_{i} \sigma_i u_i v_i^{\!\top}$ with
$\sigma_1 \geq \sigma_2 \geq \cdots \geq 0$, where $u_i \in \mathbb{R}^M$
and $v_i \in \mathbb{R}^T$ are the unit-norm train-side
(segment-indexed) and test-side singular vectors associated with the
$i$-th singular value $\sigma_i$. We define
\begin{equation}
\label{eq:ri}
  r_i \;=\; \frac{\sigma_i^2}{\lVert S^{\mathrm{seg}} \rVert_F^2}
       \;=\; \frac{\sigma_i^2}{\sum_{\ell} \sigma_\ell^2} \;\in\; [0, 1],
\end{equation}
where $\|\ldotp\|_F$ denotes Frobenius norm~\cite{golub1996matrix} and
hence $r_i$ is the fraction of Frobenius energy carried by the $i$-th
singular component. The leading ratio $r_1$ in \Cref{eq:ri} quantifies \emph{single-axis
energy concentration}. When $r_1$ is close to $1$ and $v_1$ is
near-constant across queries, every column of $S^{\mathrm{seg}}$ is
approximately a scalar multiple of the same $u_1$, so that the attributed
group is effectively query-independent. The trailing ratio
$r_{2:5} = \sum_{i=2}^{5} r_i$ summarizes how much energy the next
four axes capture beyond $u_1$. Note that $r_1$ can remain small when
query-independent structure is split across several low-rank axes of
comparable energy, a case where $\kappa$ is more discriminative because
it aggregates pairwise column similarity without assuming a single
dominant axis.

\paragraph{Mean concentration ratio.}
We quantify the tendency of a
scoring method to assign nearly the same score to
all segments within a query with the following mean
concentration ratio
\begin{equation}
\label{eq:p}
  p \;=\; \frac{1}{T}\sum_{j=1}^{T}
            \frac{M\,\mu_j^{2}}
                 {\lVert S^{\mathrm{seg}}_{\cdot j} \rVert_{2}^{2}}
        \;\in\; [0, 1],
  \qquad
  \mu_j \;=\; \frac{1}{M}\sum_{i=1}^{N}\sum_{k=1}^{n_i} S^{\mathrm{seg}}_{(i,k),\, j} \;\in\; \mathbb{R}.
\end{equation}
Each per-query term measures the fraction of total column energy
attributable to the column mean, and $p$ averages this fraction over
queries. A value close to $1$ means each column is approximately
constant across segments, so the score is explained by a per-query
offset alone. A value close to $0$ means the column mean is negligible,
so whatever signal is present resides in segment-to-segment variation
rather than a per-query offset.

All three diagnostics operate on $S^{\mathrm{seg}}$ prior to per-query
normalization and track aggregation.

% ============================================================
\subsection{Evidence Channels}
\label{sec:channels}
% ============================================================

We characterize the attributed group along a small set of \emph{evidence
channels}, each aggregating a family of features that probe a single
musical aspect. A \emph{feature} $d$ is a descriptor extracted from a 
track (e.g., a melodic-interval histogram or an MFCC statistic), taking
values in a feature-specific space $\mathcal{X}_d$, and each feature is 
paired with a feature-specific similarity function 
$\mathrm{sim}_d \colon \mathcal{X}_d \times \mathcal{X}_d \to \mathbb{R}$ 
that compares two feature values. A channel $c$ is then specified by 
its feature set $F_c$ and the corresponding similarity functions 
$\{\mathrm{sim}_d\}_{d \in F_c}$. In practice, $\mathrm{sim}_d$ is a standardized Euclidean similarity~\citep{mardia1979multivariate} for vector- and histogram-valued features, with the exception of chord progression sequences. The set of channels and the choice of features depend on the modality of
$D$, since audio data exposes timbral information that symbolic data does
not, and symbolic data permits exact melodic, harmonic, and rhythmic
descriptors that must be estimated from the audio signal. We therefore instantiate a channel set per experiment.

\paragraph{Symbolic domain experiment.}
The symbolic channel set is melody, harmony, rhythm, dynamic, and texture,
with all features computed via jSymbolic~2.2~\citep{mckay2018jsymbolic}
from the decoded MIDI. The melody channel uses the melodic-interval
histogram. The harmony channel uses pitch-class and vertical-interval
histograms. The rhythm channel uses note density, mean rhythmic value,
and rhythmic-value histograms. The dynamic channel uses inter-onset
velocity-change statistics and velocity range. The texture channel uses
polyphony statistics and pitch range.

\paragraph{Audio domain experiment.}
The audio channel set is rhythm, harmony, and timbre. The rhythm
channel combines beat-pattern features from joint beat and downbeat
tracking~\citep{heydari2021beatnet} with onset interval histograms from
librosa~\citep{mcfee2015librosa}. The harmony channel combines chroma
and Tonnetz vectors with chord-progression similarity computed via
Omnizart~\citep{wu2021omnizart}. The timbre channel uses MFCC and
constant-Q transform~\citep{schorkhuber2010constant} statistics from librosa.

Full feature lists, extraction parameters, and similarity definitions
for both experiments are given in Appendix~\ref{app:features}.

% ============================================================
\subsection{Within-Group Musical Homogeneity}
\label{sec:homogeneity}
% ============================================================
\ARIA's homogeneity analysis characterizes the attributed group
$\mathcal{A}_K(q)$, the set of $K$ training tracks with the highest
track-level scores for query $q$ under $S^{\mathrm{track}}$, by
measuring within-group pairwise similarity along each channel rather
than query--train proximity.
For each feature $d$ within channel $c$, the within-group mean
pairwise similarity for query $q$ is
\begin{equation}
\label{eq:gd}
  g_d(q) \;=\; \frac{1}{\binom{K}{2}}
    \sum_{\{x, x'\} \subseteq \mathcal{A}_K(q),\ x \neq x'}
    \mathrm{sim}_d(x, x').
\end{equation}
The per-feature null mean $\mu_d^{\mathrm{null}}$ and standard deviation
$\sigma_d^{\mathrm{null}}$ are estimated from $B = 200$ random reference
groups of size $K$ sampled uniformly without replacement from $D$, and
depend only on the training pool and group size, so they are computed
once and reused across all queries and scoring methods.
The channel-level null mean $\mu_c^{\mathrm{null}}$ and standard
deviation $\sigma_c^{\mathrm{null}}$ are obtained by computing
$\tilde{g}_c$ on the same $B$ random reference groups. The channel
z-score is
\begin{equation}
\label{eq:zc}
  z_c(q) \;=\;
    \frac{\tilde{g}_c(q) - \mu_c^{\mathrm{null}}}{\sigma_c^{\mathrm{null}}},
  \qquad
  \tilde{g}_c(q) \;=\;
    \frac{1}{|F_c|}\sum_{d \in F_c}
      \frac{g_d(q) - \mu_d^{\mathrm{null}}}{\sigma_d^{\mathrm{null}}}.
\end{equation}
Under the random reference distribution $z_c$ in \Cref{eq:zc} has zero mean and unit
variance, so positive $z_c(q)$ indicates that the attributed group is
more internally homogeneous than a random reference group of the same
size.

\paragraph{Summary metrics.}
The query-level z-score $z_c(q)$ is aggregated over all $T$ test queries
into three summary metrics per channel, where the bar in $\bar{z}_c$
denotes the across-query mean,
\begin{equation}
\label{eq:summary}
  \bar{z}_c = \frac{1}{T}\sum_{q \in Q} z_c(q), \quad
  \mathrm{Pos}_c = \frac{1}{T}\sum_{q \in Q} \mathbf{1}[z_c(q) > 0], \quad
  \mathrm{Sig}_c = \frac{1}{T}\sum_{q \in Q} \mathbf{1}[z_c(q) > 1.96].
\end{equation}
$\bar{z}_c$ reports the average effect size and $\mathrm{Pos}_c$
measures how consistently the attributed group exceeds the random
baseline. $\mathrm{Sig}_c$ reports the fraction of queries with
$z_c(q) > 1.96$, which corresponds to a one-sided $2.5\%$ test under
the random reference distribution.
% ============================================================
\section{Experiments}
\label{sec:experiments}
% ============================================================

We instantiate \ARIA in two experiments. The
first is a symbolic-music model where LDS attribution ground truth and
exact musical similarity are both available, allowing us to validate the
reliability diagnostics against an independent attribution oracle
(\Cref{sec:dattri_gt}). The second is a musical audio generation model at
a scale where LDS is computationally infeasible, where we apply \ARIA to
attribution methods and embedding-based retrieval baselines and characterize their stage-channel profiles (\Cref{sec:results_attribution,sec:results_retrieval,sec:residual_analysis}).
The two experiments use overlapping but distinct method sets. The symbolic side uses the four attribution methods from the \texttt{dattri} benchmark, while the audio side adds \FactGraSS~\citep{hu2025grass} and \LoGra~\citep{choe2024logra}, recent methods designed for the gradient and parameter scale of large generation models.

\paragraph{Symbolic domain experiment.}
We use the dattri benchmark~\citep{deng2024dattri} to obtain a pre-trained
MusicTransformer~\citep{huang2019music} on MAESTRO~\citep{hawthorne2018maestro}
with a published LDS attribution ground truth, and run \Trak~\citep{park2023trak}
(10-ensemble), TracIn~\citep{pruthi2020estimating},
\GradCos~\citep{charpiat2019input}, and GradDot~\citep{charpiat2019input}
through the benchmark's evaluation pipeline. Each score matrix is $5000 \times 178$,
with training instances indexed at the segment level so that
$S^{\mathrm{track}} = S^{\mathrm{seg}}$. We use the five jSymbolic 2.2 channels
(melody, harmony, rhythm, dynamic, texture), extracted per training segment
from the decoded MIDI. The homogeneity null is built from $B{=}200$ random
reference groups per $K \in \{20, 50, 100, 200, 300, 400, 500\}$.

\paragraph{Audio domain experiment.}
We instantiate \ARIA on a MusicLM-style~\citep{agostinelli2023musiclm,borsos2023audiolm}
three-stage hierarchical musical audio generation model trained on FMA
Large~\citep{defferrard2017fma}. The hierarchical structure provides three independently attributable
stages, semantic, coarse, and fine, each predicting a different token type
from a different audio segment duration, which allows us to measure whether
each stage's attribution recovers a different musical aspect. We adapt the
\texttt{open-musiclm}\footnote{\url{https://github.com/zhvng/open-musiclm}}
codebase with open-weight components and generate one query token sequence
per stage from each held-out evaluation track, giving $T_\ell = 7{,}148$
queries per stage. Full architecture, training, and attribution implementation
details are in Appendix~\ref{adxsec:experiments}. We refer to each (method,
stage) pair as a \emph{setting}.

% ============================================================
\subsection{Symbolic Ground-Truth Validation}
\label{sec:dattri_gt}
% ============================================================

\begin{table}[t]
  \centering
    \caption{Symbolic ground-truth validation on MusicTransformer + MAESTRO. Reliability diagnostics ($r_1$, $r_{2:5}$, $p$, $\kappa$) and within-group homogeneity $\bar z$ at $K{=}300$ across the five jSymbolic channels (M, H, R, D, T: melody, harmony, rhythm, dynamic, texture). LDS is the published attribution ground truth.}
  \label{tab:dattri_combined}
  \small
  \setlength{\tabcolsep}{4pt}
  \begin{tabular}{lccccccccccc}
  \toprule
  & & \multicolumn{4}{c}{Reliability ($S^{\mathrm{seg}}$)} & & \multicolumn{5}{c}{Homogeneity $\bar z$ at $K{=}300$} \\
  \cmidrule(lr){3-6}\cmidrule(lr){8-12}
  Method & LDS & $r_1$ & $r_{2:5}$ & $p$ & $\kappa$ & & M & H & R & D & T \\
  \midrule
  TRAK10  & \textbf{0.318} & \textbf{0.047} & \textbf{0.067} & \textbf{0.0002} & \textbf{0.022} & & $+0.28$ & $+0.38$ & $-0.25$ & $+0.13$ & $+0.18$ \\
  TracIn  & 0.149 & 0.102 & 0.229 & 0.037 & 0.106 & & $+0.95$ & $\mathbf{+2.27}$ & $+0.19$ & $+0.11$ & $+0.44$ \\
  GradCos & 0.112 & 0.137 & 0.241 & 0.038 & 0.123 & & $-0.17$ & $+0.49$ & $-0.21$ & $+0.16$ & $\mathbf{-0.14}$ \\
  GradDot & 0.089 & 0.147 & 0.238 & 0.035 & 0.129 & & $+0.55$ & $\mathbf{+1.57}$ & $+0.28$ & $\mathbf{-0.45}$ & $+0.33$ \\
  \bottomrule
  \end{tabular}
\end{table}

\Cref{tab:dattri_combined} shows that all four reliability metrics rank the four methods identically to LDS, with \Trak strongest and GradDot weakest. Method gaps are larger for the higher-order summaries ($p$ ranges over a ${\sim}150\times$ ratio between \Trak and GradDot, and $\kappa$ over ${\sim}6\times$) than for $r_1$ alone (${\sim}3\times$), consistent with the multi-axis redundancy that $r_1$ misses by construction. When reliability diagnostics indicate query-independent behavior, elevated $\bar{z}$ on any channel reflects the musical coherence of the fixed collapsed group rather than query-relevant attribution signal.

\Trak achieves the lowest $\kappa$ and $p$ among all four methods, and its moderate positive $\bar{z}$ across melody, harmony, dynamics, and texture ($+0.28$, $+0.38$, $+0.13$, $+0.18$) is accordingly interpretable as genuine query-relevant attribution. TracIn and GradDot share a common base computation, the inner product between the parameter gradient of a training example and that of a test example, with TracIn accumulating this quantity over multiple checkpoints weighted by learning rate. Both methods exhibit elevated $\kappa$ ($0.106$ and $0.129$), confirming that their score rankings are largely query-independent, and the anomalously high harmonic $\bar{z}$ at $K{=}300$ ($+2.27$ and $+1.57$) is therefore a symptom of this collapse rather than evidence of meaningful harmonic attribution. GradCos normalizes each gradient by its $\ell_2$ norm, replacing inner product with cosine similarity. This removes the harmonic dominance that drives collapse in TracIn and GradDot, lowering harmonic $\bar{z}$ to $+0.49$ and yielding a channel profile distinct from those methods. The collapse itself persists, however, with $\kappa{=}0.123$ comparable to TracIn and GradDot and the attributed group remaining query-independent across all four channels. Magnitude normalization changes the channel profile of collapse but does not restore query-dependent attribution. Full $K$-sweep results for all settings are provided in Appendix~\ref{app:full_result}.

% ============================================================
\subsection{Audio Attribution Method Results}
\label{sec:results_attribution}
% ============================================================

\begin{table}[t]
\centering
\caption{Reliability diagnostics and within-group homogeneity for all 15 settings at $K{=}300$ ($N_{\text{test}}{=}7{,}148$). Stage: S/C/F = semantic/coarse/fine. Bold marks positive $\bar z$ and the best (lowest) reliability value within each method group.}
\label{tab:attribution_coherence}
\vskip 0.1in
\resizebox{\linewidth}{!}{%
\begin{tabular}{llcccc rrr rrr rrr}
\toprule
& & & & & & \multicolumn{3}{c}{Rhythmic} & \multicolumn{3}{c}{Harmonic} & \multicolumn{3}{c}{Timbral} \\
\cmidrule(lr){7-9}\cmidrule(lr){10-12}\cmidrule(lr){13-15}
Method & Stage & $r_1$ & $r_{2:5}$ & $p$ & $\kappa$
  & $\bar{z}$ & Pos$\uparrow$ & Sig$\uparrow$
  & $\bar{z}$ & Pos$\uparrow$ & Sig$\uparrow$
  & $\bar{z}$ & Pos$\uparrow$ & Sig$\uparrow$ \\
\midrule
\multirow{3}{*}{\Trak}
  & S & \textbf{0.282} & 0.098 & 0.175 & \textbf{0.076}
    & \textbf{+1.51} & 85.8\% & 38.1\%
    & $-$3.63        &  1.5\% &  0.0\%
    & $-$0.86        & 28.3\% &  3.3\% \\
  & C & 0.993 & 0.004 & 0.022 & 0.991
    & $-$2.78        &  0.1\% & 0.0\%
    & $-$5.89        &  0.0\% & 0.0\%
    & $-$7.58        &  0.0\% & 0.0\% \\
  & F & 0.772 & 0.208 & \textbf{0.002} & 0.641
    & \textbf{+0.45} & 76.1\% &  0.4\%
    & \textbf{+1.25} & 93.9\% & 17.5\%
    & \textbf{+0.73} & 89.1\% &  1.6\% \\
\midrule
\multirow{3}{*}{\FactGraSS}
  & S & \textbf{0.006} & 0.018 & \textbf{0.000} & \textbf{0.012}
    & \textbf{+1.08} & 61.2\% & 40.2\%
    & $-$0.83        & 39.3\% & 22.6\%
    & \textbf{+2.47} & 64.5\% & 51.5\% \\
  & C & 0.051 & 0.030 & 0.001 & 0.046
    & $-$0.57        & 36.9\% & 14.0\%
    & $-$3.45        & 13.2\% &  5.6\%
    & $-$1.60        & 32.6\% & 21.6\% \\
  & F & 0.866 & 0.081 & \textbf{0.000} & 0.663
    & $-$0.80        & 15.2\% &  0.0\%
    & $-$3.57        &  0.0\% &  0.0\%
    & $-$4.81        &  0.0\% &  0.0\% \\
\midrule
\multirow{3}{*}{\GradCos}
  & S & \textbf{0.413} & 0.419 & 0.216 & \textbf{0.374}
    & \textbf{+0.36} & 57.6\% & 18.6\%
    & $-$2.15        & 21.7\% &  5.7\%
    & \textbf{+0.93} & 67.1\% & 30.5\% \\
  & C & 1.000 & 0.000 & 1.000 & 0.997
    & $-$2.81        & 14.2\% &  5.5\%
    & \textbf{+8.56} & 99.2\% & 97.6\%
    & \textbf{+29.56}& 99.8\% & 99.7\% \\
  & F & 0.793 & 0.200 & \textbf{0.015} & 0.736
    & $-$0.35        & 28.5\% &  0.4\%
    & $-$5.54        &  0.0\% &  0.0\%
    & $-$10.53       &  0.7\% &  0.1\% \\
\midrule
\multirow{3}{*}{\LoGra}
  & S & \textbf{0.008} & 0.017 & \textbf{0.000} & \textbf{0.013}
    & \textbf{+0.96} & 61.3\% & 36.8\%
    & $-$1.53        & 30.7\% & 15.1\%
    & \textbf{+1.41} & 59.7\% & 44.4\% \\
  & C & 0.057 & 0.038 & 0.006 & 0.049
    & $-$0.38        & 42.7\% & 21.7\%
    & $-$3.04        & 23.2\% & 13.4\%
    & \textbf{+1.23} & 52.1\% & 39.6\% \\
  & F & 0.223 & 0.065 & \textbf{0.000} & 0.272
    & $-$0.77        & 22.4\% &  0.6\%
    & $-$3.90        &  0.0\% &  0.0\%
    & $-$4.45        &  0.0\% &  0.0\% \\
\midrule
CLAP  & --- & 0.924 & 0.054 & 0.801 & 0.584
    & \textbf{+1.38} & 60.9\% & 39.8\%
    & $-$2.75        & 30.0\% & 17.2\%
    & \textbf{+1.50} & 59.3\% & 48.2\% \\
CLEWS & --- & 0.607 & 0.141 & 0.406 & \textbf{0.280}
    & \textbf{+0.50} & 51.0\% & 26.3\%
    & $-$3.04        & 34.9\% & 24.8\%
    & \textbf{+0.71} & 54.4\% & 38.7\% \\
MERT  & --- & \textbf{0.437} & 0.375 & \textbf{0.002} & 0.358
    & \textbf{+1.09} & 60.4\% & 39.5\%
    & \textbf{+6.55} & 86.8\% & 78.2\%
    & \textbf{+4.84} & 84.5\% & 74.7\% \\
\bottomrule
\end{tabular}%
}
\vskip -0.1in
\end{table}

\Cref{tab:attribution_coherence} reports reliability diagnostics
($r_1$, $r_{2:5}$, $p$, $\kappa$) and within-group homogeneity ($\bar{z}_c$, Pos, Sig) for all fifteen settings at $K{=}300$. We read homogeneity results alongside the reliability diagnostics: high $r_1$ or $p$ indicates a collapsed matrix that retrieves the same fixed group for every query, making its z-scores uninformative about attribution quality.

\paragraph{Collapsed settings.}
Several settings exhibit high $r_1$, meaning the score matrix is
dominated by a single rank-1 component and every query retrieves the same static group.
In these settings, homogeneity z-scores reflect the within-group musical homogeneity
of that fixed set, not query-specific attribution.
\GradCos coarse is the most extreme case ($r_1{=}1.000$, $p{=}1.000$),
producing anomalously large harmonic and timbral z-scores
($+8.56$ and $+29.56$) that reflect the homogeneity of the fixed static group.
\Trak coarse ($r_1{=}0.993$) and \FactGraSS fine ($r_1{=}0.866$) are similarly
collapsed and yield uniformly negative z-scores, indicating that their fixed
attributed groups are less musically homogeneous than random groups of the same size.

\paragraph{Query-dependent settings.}
Settings with low $r_1$ and low $p$ retrieve query-specific groups,
so their z-scores reflect genuine attribution-driven musical homogeneity.
\FactGraSS semantic produces the strongest positive timbral z-score among attribution settings
($+2.47$, $\mathrm{Sig}{=}51.5\%$) along with positive rhythmic ($+1.08$).
\LoGra semantic ($+1.41$) and \GradCos semantic ($+0.93$) also yield
positive timbral homogeneity, and \LoGra coarse produces positive
timbral homogeneity ($+1.23$).
\Trak semantic ($r_1{=}0.282$, $p{=}0.175$) produces a positive rhythmic
z-score ($+1.51$, $\mathrm{Pos}{=}85.8\%$).
Among query-dependent settings, harmonic $\bar{z}$ remains negative for
every \FactGraSS, \GradCos, and \LoGra setting at every stage.
\Trak fine ($r_1{=}0.772$, $p{=}0.002$) is the only attribution setting
with positive $\bar{z}$ on all three channels ($+0.45$, $+1.25$, $+0.73$).
Since \Trak fine is the only non-collapsed attribution setting with positive harmonic
homogeneity, we additionally verify in
Appendix~\ref{app:genre_confound} that this z-score is not driven by
genre-cluster confounds in the random reference distribution.

% ============================================================
\subsection{Embedding Retrieval Method Results}
\label{sec:results_retrieval}
% ============================================================

\begin{table}[h]
  \centering
  \caption{Axis-channel alignment of $u_1$ for embedding retrieval baselines. $\alpha_c^{\max}$: single-dimension max correlation. $\alpha_c^{\mathrm{reg}}$: multivariate OLS $\sqrt{R^2}$. Bold marks the best value in each column.}
  \label{tab:alpha_baselines}
  \small
  \begin{tabular}{lcccccc}
  \toprule
  & \multicolumn{2}{c}{Rhythmic} & \multicolumn{2}{c}{Harmonic} & \multicolumn{2}{c}{Timbral} \\
  \cmidrule(lr){2-3}\cmidrule(lr){4-5}\cmidrule(lr){6-7}
  Method & $\alpha^{\max}$ & $\alpha^{\mathrm{reg}}$ & $\alpha^{\max}$ & $\alpha^{\mathrm{reg}}$ & $\alpha^{\max}$ & $\alpha^{\mathrm{reg}}$ \\
  \midrule
  CLAP  & 0.186 & 0.243 & 0.306 & 0.361 & 0.245 & 0.493 \\
  CLEWS & 0.117 & 0.143 & 0.143 & 0.233 & 0.333 & 0.453 \\
  MERT  & \textbf{0.231} & \textbf{0.316} & \textbf{0.482} & \textbf{0.680} & \textbf{0.780} & \textbf{0.896} \\
  \bottomrule
  \end{tabular}
\end{table}

\paragraph{Retrieval baselines.}
\CLAP ($p{=}0.801$) is dominated by a query-independent offset, producing negative harmonic z-scores ($-2.75$) despite positive rhythmic and timbral results. \CLEWS ($p{=}0.406$) shows intermediate offset behavior with similarly negative harmonic z-scores ($-3.04$). \MERT has negligible mean offset ($p{=}0.002$) and moderate rank-1 concentration ($r_1{=}0.437$), making it query-dependent. It produces the strongest harmonic ($+6.55$) and timbral ($+4.84$) homogeneity of any non-collapsed setting.

The three retrieval baselines share the same scoring rule yet produce substantially different channel-specific homogeneity profiles in \Cref{tab:attribution_coherence}. The differences therefore originate from how each encoder maps onto musical channels. We decompose this difference along two measurable structural quantities, the alignment between each encoder's dominant retrieval axis and the musical channels.

\paragraph{Axis-channel alignment.}
To characterize what each encoder's dominant retrieval axis captures,
we measure its alignment with per-track musical features across
channels. The dominant left singular vector $u_1 \in \mathbb{R}^N$ of
$S^{\mathrm{track}}$ characterizes each encoder's primary ranking
direction. We quantify the alignment between $u_1$ and channel $c$
with two complementary measures:
\begin{equation}
\label{eq:alpha}
  \alpha_c^{\max} \;=\; \max_{d \in F_c}
    \lvert \mathrm{corr}(u_1, \psi_d) \rvert,
  \qquad
  \alpha_c^{\mathrm{reg}} \;=\; \max_{\Psi}
    \sqrt{R^2(u_1 \mid \Psi)},
\end{equation}
where $\psi_d \in \mathbb{R}^N$ is the per-track value of feature $d$
and $\Psi \in \mathbb{R}^{N \times D}$ ranges over the per-track
feature matrices of the feature groups in channel $c$ (e.g.\ the
12-dimensional chroma vector).
$\alpha_c^{\max}$ in \Cref{eq:alpha} detects alignment concentrated in a single feature
dimension, while $\alpha_c^{\mathrm{reg}}$ captures alignment
distributed across a feature group.
A high value on either indicates that training tracks ranking high on
$u_1$ tend to be musically similar along channel $c$, revealing which
musical aspect the encoder's dominant axis encodes.

The three encoders differ sharply in which channels their dominant
axis aligns with (\Cref{tab:alpha_baselines}), and these differences
reflect each encoder's pretraining objective. \MERT, which
self-supervised pretraining uses a Constant-Q Transform based musical
teacher and a much larger music corpus, achieves both the strongest
harmonic alignment ($\alpha_c^{\mathrm{reg}} = 0.680$ on the harmonic
channel) and near-complete timbral alignment ($0.896$). \CLEWS is
trained for cover song detection, a task that rewards a holistic
match of musical identity across heavy stylistic variation rather
than agreement on any single feature, which plausibly explains why
its alignment is weak across all three channels and lowest among the
three on harmonic. \CLAP is trained with a contrastive objective
that pairs audio and text with surface-level descriptions of sound sources,
which biases its representation toward broad sonic categories rather
than fine-grained musical aspect and yields intermediate alignment
without strong specialization toward any single channel. Embedding-
based retrieval thus tells us what each encoder is sensitive to, not
the influence of training data on the generative model that produced.

% ============================================================
\subsection{Residual Analysis}
\label{sec:residual_analysis}
% ============================================================

\begin{figure}[t]
  \centering
  \begin{subfigure}[t]{0.245\textwidth}
    \includegraphics[width=\linewidth]{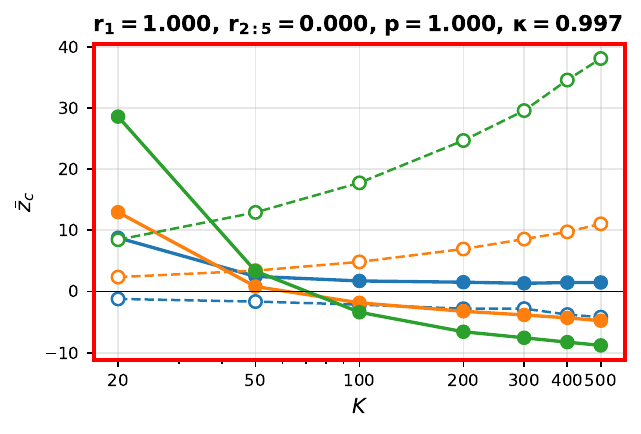}
    \caption{\GradCos coarse}\label{fig:ksweep_main:gradcos_coarse}
  \end{subfigure}\hfill
  \begin{subfigure}[t]{0.245\textwidth}
    \includegraphics[width=\linewidth]{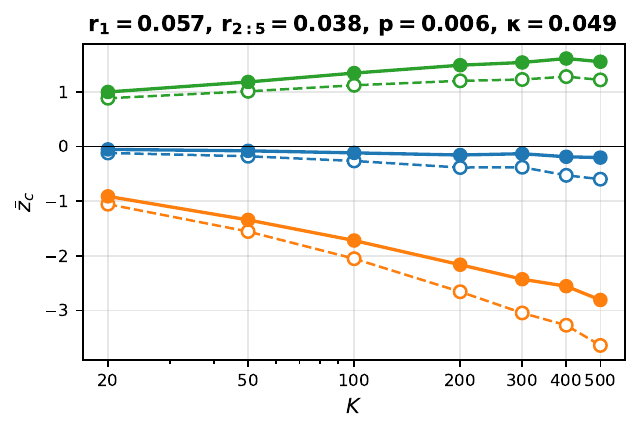}
    \caption{\LoGra coarse}\label{fig:ksweep_main:logra_coarse}
  \end{subfigure}\hfill
  \begin{subfigure}[t]{0.245\textwidth}
    \includegraphics[width=\linewidth]{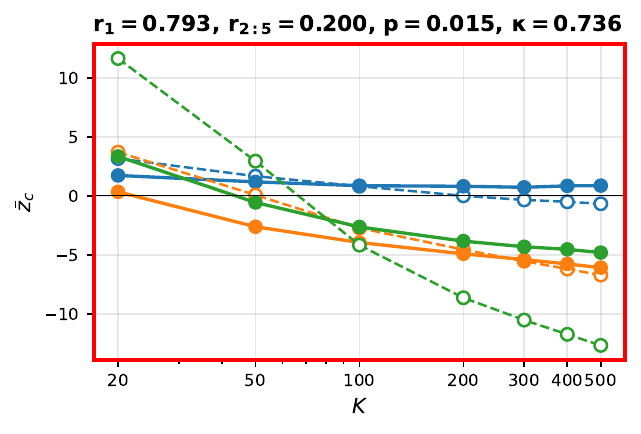}
    \caption{\GradCos fine}\label{fig:ksweep_main:gradcos_fine}
  \end{subfigure}\hfill
  \begin{subfigure}[t]{0.245\textwidth}
    \includegraphics[width=\linewidth]{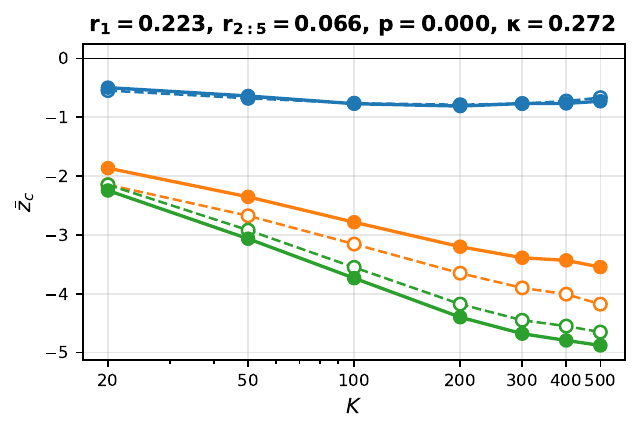}
    \caption{\LoGra fine}\label{fig:ksweep_main:logra_fine}
  \end{subfigure}\\[0.3em]
  \includegraphics[width=0.85\textwidth]{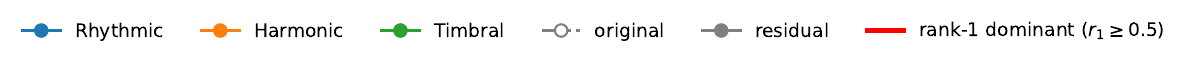}
  \caption{Homogeneity $\bar z_c$ across $K$ for \GradCos and \LoGra at coarse and fine, contrasting the largest and smallest $r_1$ regimes among non-semantic attribution settings. Dashed curves use the original $S^{\mathrm{seg}}$, solid curves use the rank-1 residual.}
  \label{fig:ksweep_main}
\end{figure}

To separate query-dependent attribution signal from rank-1 axis bias, we subtract $\sigma_1 u_1 v_1^\top$ from $S^{\mathrm{seg}}$ and reapply the homogeneity analysis to the residual. \Cref{fig:ksweep_main} shows \GradCos and \LoGra at coarse and fine, the four settings that span the largest and smallest $r_1$ values among non-semantic attribution settings. The remaining settings are reported in Appendix~\ref{app:full_result}.

\paragraph{Rank-1 removal.}
A setting with $r_1$ near $1$ and $r_{2:5}$ near $0$ has a rank-one $S^{\mathrm{seg}}$, so its $\bar z_c$ measures the within-group homogeneity of the fixed $u_1$ group rather than any query-specific attribution. \GradCos coarse is the clearest case ($r_1{=}1.000$, $r_{2:5}{=}0.000$), producing $\bar z_c{=}{+}29.56$ on timbral and $+8.56$ on harmonic, both the largest values across all $15$ settings. Rank-1 removal flips them to $-7.57$ and $-3.84$, confirming that the original signal is the static $u_1$ group rather than an attribution result. \GradCos fine ($r_1{=}0.793$, $r_{2:5}{=}0.200$) is also rank-1 dominant but carries some residual signal beyond $u_1$, so the gap between original and residual is smaller than at coarse. \LoGra ($r_1{=}0.057$ at coarse, $0.223$ at fine) places only a small fraction of its Frobenius energy on $u_1$, so the residual $\bar z_c$ stays close to the original by construction.

\paragraph{Implication for method comparison.}
A naive ranking on $\bar z_c$ would place \GradCos coarse at the top, since its rank-one $S^{\mathrm{seg}}$ produces the largest homogeneity number in our experiment. The $(r_1, r_{2:5})$ check identifies these settings as carrying no query-dependent signal, so they cannot be compared to others on homogeneity and must be excluded from method ranking rather than counted as strong attribution evidence.
\FloatBarrier
\section{Conclusions and Limitations}
\label{sec:discussion}
We propose \ARIA, a framework that decomposes attribution into musical aspects and diagnoses score-matrix quality without ground truth.
This addresses two structural limits of attribution evaluation in music generation, since per-aspect ground truth is hard to construct and counterfactual retraining at the scale of musical audio generation models is computationally infeasible. These limits matter because, with music copyright lawsuits actively underway~\citep{umg2024suno, umg2024udio}, attribution evidence linking model outputs to specific training data has become increasingly important, yet a single scalar cannot support such evidence under the courts' idea-expression distinction~\citep{mullensiefen2009court, dornis2025copyright}.

On the symbolic music benchmark, all four reliability diagnostics rank attribution methods identically to LDS, supporting their use as a substitute signal where LDS is intractable. Applying \ARIA to a MusicLM-style musical audio generation model, the reliability diagnostics exclude the settings that would otherwise top a naive $\bar z_c$ ranking, with rank-1 residual analysis reversing the sign of their homogeneity and confirming that the original signal is not attribution evidence. Embedding retrieval baselines diverge along the lines of each encoder's pretraining objective, indicating that they report what the encoder represents rather than the influence of training data on the generative model. \ARIA diagnoses the structure of an attribution score matrix and the within-group homogeneity of its top-$K$ retrievals; establishing causal influence at the per-track level remains future work, addressable through controlled synthetic data. 

By reframing music attribution as a multi-aspect, diagnosis-first problem, \ARIA provides a foundation on which the field can build richer evidentiary tools as music generation models continue to scale, bringing principled attribution closer to the compensation and copyright frameworks the field increasingly needs.

\clearpage
{\small
\bibliographystyle{plain}
\bibliography{refs}

\begin{thebibliography}{10}

\bibitem{agostinelli2023musiclm}
Andrea Agostinelli, Timo~I Denk, Zal{\'a}n Borsos, Jesse Engel, Mauro Verzetti,
  Antoine Caillon, Qingqing Huang, Aren Jansen, Adam Roberts, Marco
  Tagliasacchi, et~al.
\newblock {MusicLM}: Generating music from text.
\newblock {\em arXiv preprint arXiv:2301.11325}, 2023.

\bibitem{akyurek-etal-2022-towards}
Ekin Akyurek, Tolga Bolukbasi, Frederick Liu, Binbin Xiong, Ian Tenney, Jacob
  Andreas, and Kelvin Guu.
\newblock Towards tracing knowledge in language models back to the training
  data.
\newblock In Yoav Goldberg, Zornitsa Kozareva, and Yue Zhang, editors, {\em
  Findings of the Association for Computational Linguistics: EMNLP 2022}, pages
  2429--2446, Abu Dhabi, United Arab Emirates, December 2022. Association for
  Computational Linguistics.

\bibitem{barnett2024exploring}
Julia Barnett, Hugo~Flores Garcia, and Bryan Pardo.
\newblock Exploring musical roots: Applying audio embeddings to empower
  influence attribution for a generative music model.
\newblock In {\em Proceedings of the 25th International Society for Music
  Information Retrieval Conference (ISMIR)}, 2024.

\bibitem{bittner2017deep}
Rachel~M. Bittner, Brian McFee, Justin Salamon, Peter Li, and Juan~Pablo Bello.
\newblock Deep salience representations for {F0} estimation in polyphonic
  music.
\newblock In {\em Proceedings of the 18th International Society for Music
  Information Retrieval Conference (ISMIR)}, pages 63--70, 2017.

\bibitem{borsos2023audiolm}
Zal{\'a}n Borsos, Rapha{\"e}l Marinier, Damien Vincent, Eugene Kharitonov,
  Olivier Pietquin, Matt Sharifi, Dominik Roblek, Olivier Teboul, David
  Grangier, Marco Tagliasacchi, and Neil Zeghidour.
\newblock {AudioLM}: A language modeling approach to audio generation.
\newblock {\em IEEE/ACM Transactions on Audio, Speech, and Language
  Processing}, 31:2523--2533, 2023.

\bibitem{carlini2023quantifying}
Nicholas Carlini, Daphne Ippolito, Matthew Jagielski, Katherine Lee, Florian
  Tram{\`e}r, and Chiyuan Zhang.
\newblock Quantifying memorization across neural language models.
\newblock In {\em International Conference on Learning Representations (ICLR)},
  2023.

\bibitem{carlini2021extracting}
Nicholas Carlini, Florian Tramer, Eric Wallace, Matthew Jagielski, Ariel
  Herbert-Voss, Katherine Lee, Adam Roberts, Tom Brown, Dawn Song, {\'U}lfar
  Erlingsson, Alina Oprea, and Colin Raffel.
\newblock Extracting training data from large language models.
\newblock In {\em 30th USENIX Security Symposium (USENIX Security 21)}, pages
  2633--2650, 2021.

\bibitem{carlini2023extracting}
Nicolas Carlini, Jamie Hayes, Milad Nasr, Matthew Jagielski, Vikash Sehwag,
  Florian Tramer, Borja Balle, Daphne Ippolito, and Eric Wallace.
\newblock Extracting training data from diffusion models.
\newblock In {\em 32nd USENIX Security Symposium (USENIX Security 23)}, pages
  5253--5270, 2023.

\bibitem{charpiat2019input}
Guillaume Charpiat, Nicolas Girard, Loris Felardos, and Yuliya Tarabalka.
\newblock Input similarity from the neural network perspective.
\newblock {\em Advances in Neural Information Processing Systems}, 32, 2019.

\bibitem{choe2024logra}
Sang~Keun Choe, Hwijeen Ahn, Juhan Bae, Kewen Zhao, Youngseog Chung, Adithya
  Pratapa, Willie Neiswanger, Emma Strubell, Teruko Mitamura, Jeff Schneider,
  Eduard Hovy, Roger~Baker Grosse, and Eric~P. Xing.
\newblock What is your data worth to {GPT}? {LLM}-scale data valuation with
  influence functions.
\newblock In {\em The Thirty-ninth Annual Conference on Neural Information
  Processing Systems}, 2026.

\bibitem{choi2025large}
Woosung Choi, Junghyun Koo, Kin~Wai Cheuk, Joan Serr{\`a}, Marco~A
  Mart{\'\i}nez-Ram{\'\i}rez, Yukara Ikemiya, Naoki Murata, Yuhta Takida,
  Wei-Hsiang Liao, and Yuki Mitsufuji.
\newblock Large-scale training data attribution for music generative models via
  unlearning.
\newblock {\em arXiv preprint arXiv:2506.18312}, 2025.

\bibitem{davis1980mfcc}
Steven Davis and Paul Mermelstein.
\newblock Comparison of parametric representations for monosyllabic word
  recognition in continuously spoken sentences.
\newblock {\em IEEE Transactions on Acoustics, Speech, and Signal Processing},
  28(4):357--366, 1980.

\bibitem{defferrard2017fma}
Micha{\"e}l Defferrard, Kirell Benzi, Pierre Vandergheynst, and Xavier Bresson.
\newblock {FMA}: A dataset for music analysis.
\newblock In {\em Proceedings of the 18th International Society for Music
  Information Retrieval Conference}, pages 316--323, 2017.

\bibitem{deng2023computational}
Junwei Deng, Xirui Jiang, Shiyuan Zhang, Shichang Zhang, Himabindu Lakkaraju,
  Ruijiang Gao, Chris Donahue, and Jiaqi~W. Ma.
\newblock Computational copyright: Towards a royalty model for music generative
  {AI}.
\newblock {\em arXiv preprint arXiv:2312.06646}, 2023.

\bibitem{deng2024dattri}
Junwei Deng, Ting-Wei Li, Shiyuan Zhang, Shixuan Liu, Yijun Pan, Hao Huang,
  Xinhe Wang, Pingbang Hu, Xingjian Zhang, and Jiaqi Ma.
\newblock dattri: A library for efficient data attribution.
\newblock In A.~Globerson, L.~Mackey, D.~Belgrave, A.~Fan, U.~Paquet,
  J.~Tomczak, and C.~Zhang, editors, {\em Advances in Neural Information
  Processing Systems}, volume~37, pages 136763--136781. Curran Associates,
  Inc., 2024.

\bibitem{dornis2025copyright}
Tim~W. Dornis and Sebastian Stober.
\newblock Generative {AI} training and copyright law.
\newblock {\em Transactions of the International Society for Music Information
  Retrieval}, 2025.
\newblock arXiv:2502.15858.

\bibitem{elizalde2023clap}
Benjamin Elizalde, Soham Deshmukh, Mahmoud Al~Ismail, and Huaming Wang.
\newblock {CLAP}: Learning audio concepts from natural language supervision.
\newblock In {\em ICASSP 2023 -- IEEE International Conference on Acoustics,
  Speech and Signal Processing}, pages 1--5, 2023.

\bibitem{esling2021flowsynth}
Philippe Esling, Naotake Masuda, and Axel Chemla-Romeu-Santos.
\newblock Flowsynth: simplifying complex audio generation through explorable
  latent spaces with normalizing flows.
\newblock In {\em Proceedings of the Twenty-Ninth International Conference on
  International Joint Conferences on Artificial Intelligence}, pages
  5273--5275, 2021.

\bibitem{golub1996matrix}
Gene~H Golub and Charles~F Van~Loan.
\newblock Matrix computations 3rd edition.
\newblock {\em The John Hopkins University, Baltimore}, 1996.

\bibitem{harte2006detecting}
Christopher Harte, Mark Sandler, and Martin Gasser.
\newblock Detecting harmonic change in musical audio.
\newblock In {\em Proceedings of the 1st ACM Workshop on Audio and Music
  Computing Multimedia}, pages 21--26, 2006.

\bibitem{hawthorne2018maestro}
Curtis Hawthorne, Andriy Stasyuk, Adam Roberts, Ian Simon, Cheng-Zhi~Anna
  Huang, Sander Dieleman, Erich Elsen, Jesse Engel, and Douglas Eck.
\newblock Enabling factorized piano music modeling and generation with the
  {MAESTRO} dataset.
\newblock In {\em International Conference on Learning Representations}, 2019.

\bibitem{herremans2017functional}
Dorien Herremans, Ching-Hua Chuan, and Elaine Chew.
\newblock A functional taxonomy of music generation systems.
\newblock {\em ACM Computing Surveys (CSUR)}, 50(5):1--30, 2017.

\bibitem{heydari2021beatnet}
Mojtaba Heydari, Frank Cwitkowitz, and Zhiyao Duan.
\newblock Beatnet: Crnn and particle filtering for online joint beat downbeat
  and meter tracking.
\newblock 2021.

\bibitem{hu2025grass}
Pingbang Hu, Joseph Melkonian, Weijing Tang, Han Zhao, and Jiaqi~W. Ma.
\newblock {GraSS}: Scalable data attribution with gradient sparsification and
  sparse projection.
\newblock {\em arXiv preprint arXiv:2505.18976}, 2025.

\bibitem{huang2019music}
Cheng-Zhi~Anna Huang, Ashish Vaswani, Jakob Uszkoreit, Ian Simon, Curtis
  Hawthorne, Noam Shazeer, Andrew~M Dai, Matthew~D Hoffman, Monica Dinculescu,
  and Douglas Eck.
\newblock Music transformer: Generating music with long-term structure.
\newblock In {\em International Conference on Learning Representations}, 2019.

\bibitem{ilyas2022datamodels}
Andrew Ilyas, Sung~Min Park, Logan Engstrom, Guillaume Leclerc, and Aleksander
  Madry.
\newblock Datamodels: Predicting predictions from training data.
\newblock In {\em ICML}, 2022.

\bibitem{kim2025noencore}
Jinju Kim, Taehan Kim, Abdul Waheed, Jong~Hwan Ko, and Rita Singh.
\newblock No encore: Unlearning as opt-out in music generation.
\newblock In {\em {NeurIPS} 2025 Workshop on {AI} for Music}, 2025.

\bibitem{koh2017understanding}
Pang~Wei Koh and Percy Liang.
\newblock Understanding black-box predictions via influence functions.
\newblock In {\em Proceedings of the 34th International Conference on Machine
  Learning}, volume~70 of {\em ICML '17}, pages 1885--1894, 2017.

\bibitem{lee2020disentangled}
Jongpil Lee, Nicholas~J Bryan, Justin Salamon, Zeyu Jin, and Juhan Nam.
\newblock Disentangled multidimensional metric learning for music similarity.
\newblock In {\em ICASSP 2020-2020 IEEE International Conference on Acoustics,
  Speech and Signal Processing (ICASSP)}, pages 6--10. IEEE, 2020.

\bibitem{lee2020metric}
Jongpil Lee, Nicholas~J Bryan, Justin Salamon, Zeyu Jin, and Juhan Nam.
\newblock Metric learning vs classification for disentangled music
  representation learning.
\newblock In {\em The 21th International Society for Music Information
  Retrieval Conference (ISMIR)}. International Society for Music Information
  Retrieval, 2020.

\bibitem{lee1988thirteen}
Joseph Lee~Rodgers and W~Alan Nicewander.
\newblock Thirteen ways to look at the correlation coefficient.
\newblock {\em The American Statistician}, 42(1):59--66, 1988.

\bibitem{li2023mert}
Yizhi Li, Ruibin Yuan, Ge~Zhang, Yinghao Ma, Xingran Chen, Hanzhi Yin, Chenghua
  Lin, Anton Ragni, Emmanouil Benetos, Norbert Gyenge, et~al.
\newblock {MERT}: Acoustic music understanding model with large-scale
  self-supervised training.
\newblock {\em arXiv preprint arXiv:2306.00107}, 2023.

\bibitem{livingston2013copyright}
Margit Livingston and Joseph Urbinato.
\newblock Copyright infringement of music: Determining whether what sounds
  alike is alike.
\newblock {\em Vanderbilt Journal of Entertainment and Technology Law},
  15(2):227--294, 2013.

\bibitem{luo2019learning}
Yin-Jyun Luo, Kat Agres, and Dorien Herremans.
\newblock Learning disentangled representations of timbre and pitch for musical
  instrument sounds using {Gaussian} mixture variational autoencoders.
\newblock In {\em Proceedings of the 20th International Society for Music
  Information Retrieval Conference (ISMIR)}, 2019.

\bibitem{mardia1979multivariate}
K.~V. Mardia, J.~T. Kent, and J.~M. Bibby.
\newblock {\em Multivariate Analysis}.
\newblock Academic Press, London, 1979.

\bibitem{mcfee2015librosa}
Brian McFee, Colin Raffel, Dawen Liang, Daniel P.~W. Ellis, Matt McVicar, Eric
  Battenberg, and Oriol Nieto.
\newblock librosa: Audio and music signal analysis in {Python}.
\newblock In {\em Proceedings of the 14th Python in Science Conference}, pages
  18--25, 2015.

\bibitem{mckay2018jsymbolic}
Cory McKay, Julie Cumming, and Ichiro Fujinaga.
\newblock {jSymbolic} 2.2: Extracting features from symbolic music for use in
  musicological and {MIR} research.
\newblock {\em Proceedings of the International Society for Music Information
  Retrieval Conference (ISMIR)}, pages 348--354, 2018.

\bibitem{morreale2025attribution}
Fabio Morreale, Wiebke Hutiri, Joan Serr{\`a}, Alice Xiang, and Yuki Mitsufuji.
\newblock Attribution-by-design: Ensuring inference-time provenance in
  generative music systems.
\newblock {\em arXiv preprint arXiv:2510.08062}, 2025.

\bibitem{mullensiefen2009court}
Daniel M{\"u}llensiefen and Marc Pendzich.
\newblock Court decisions on music plagiarism and the predictive value of
  similarity algorithms.
\newblock {\em Musicae Scientiae}, 13(1\_suppl):257--295, 2009.

\bibitem{muller2015fundamentals}
Meinard M{\"u}ller.
\newblock {\em Fundamentals of Music Processing}.
\newblock Springer, 2015.

\bibitem{nicolas2023harmonizing}
Peter Nicolas.
\newblock Harmonizing music theory and music law.
\newblock {\em Iowa Law Review}, 108:1247--1313, 2023.

\bibitem{theverge2026campbell}
Terrence O'Brien.
\newblock A folk musician became a target for {AI} fakes and a copyright troll.
\newblock The Verge, April 2026.

\bibitem{park2023trak}
Sung~Min Park, Kristian Georgiev, Andrew Ilyas, Guillaume Leclerc, and
  Aleksander Madry.
\newblock {TRAK}: Attributing model behavior at scale.
\newblock In {\em Proceedings of the 40th International Conference on Machine
  Learning}, pages 27074--27113, 2023.

\bibitem{park2025concepttrak}
Yonghyun Park, Chieh-Hsin Lai, Satoshi Hayakawa, Yuhta Takida, Naoki Murata,
  Wei-Hsiang Liao, Woosung Choi, Kin~Wai Cheuk, Junghyun Koo, and Yuki
  Mitsufuji.
\newblock {Concept-TRAK}: Understanding how diffusion models learn concepts
  through concept-level attribution.
\newblock In {\em International Conference on Learning Representations (ICLR)},
  2026.

\bibitem{pruthi2020estimating}
Garima Pruthi, Frederick Liu, Satyen Kale, and Mukund Sundararajan.
\newblock Estimating training data influence by tracing gradient descent.
\newblock In {\em Advances in Neural Information Processing Systems},
  volume~33, pages 19920--19930, 2020.

\bibitem{salamon2014melody}
Justin Salamon, Emilia G{\'o}mez, Daniel P.~W. Ellis, and Ga{\"e}l Richard.
\newblock Melody extraction from polyphonic music signals: Approaches,
  applications, and challenges.
\newblock {\em IEEE Signal Processing Magazine}, 31(2):118--134, 2014.

\bibitem{schorkhuber2010constant}
Christian Sch{\"o}rkhuber and Anssi Klapuri.
\newblock Constant-q transform toolbox for music processing.
\newblock In {\em 7th sound and music computing conference, Barcelona, Spain},
  pages 3--64. SMC, 2010.

\bibitem{serra2025clews}
Joan Serr{\`a}, R.\~Oguz Araz, Dmitry Bogdanov, and Yuki Mitsufuji.
\newblock Supervised contrastive learning from weakly-labeled audio segments
  for musical version matching.
\newblock In {\em International Conference on Machine Learning ({ICML})}, 2025.

\bibitem{somepalli2023diffusion}
Gowthami Somepalli, Vasu Singla, Micah Goldblum, Jonas Geiping, and Tom
  Goldstein.
\newblock Diffusion art or digital forgery? investigating data replication in
  diffusion models.
\newblock In {\em Proceedings of the IEEE/CVF conference on computer vision and
  pattern recognition}, pages 6048--6058, 2023.

\bibitem{tatar2021latent}
K{\i}van{\c{c}} Tatar, Daniel Bisig, and Philippe Pasquier.
\newblock Latent timbre synthesis: Audio-based variational auto-encoders for
  music composition and sound design applications.
\newblock {\em Neural Computing and Applications}, 33(1):67--84, 2021.

\bibitem{umg2024suno}
{UMG Recordings v. Suno}.
\newblock Complaint, {UMG Recordings, Inc.} v. {Suno, Inc.}, no. 1:24-cv-11611
  ({D. Mass. 2024}), 2024.

\bibitem{umg2024udio}
{UMG Recordings v. Udio}.
\newblock Complaint, {UMG Recordings, Inc.} v. {Uncharted Labs, Inc.}, no.
  1:24-cv-04777 ({S.D.N.Y. 2024}), 2024.

\bibitem{wang2023evaluating}
Sheng-Yu Wang, Alexei~A. Efros, Jun-Yan Zhu, and Richard Zhang.
\newblock Evaluating data attribution for text-to-image models.
\newblock In {\em Proceedings of the IEEE/CVF International Conference on
  Computer Vision (ICCV)}, pages 7192--7203, 2023.

\bibitem{wang2024unlearning}
Sheng-Yu Wang, Aaron Hertzmann, Alexei~A Efros, Jun-Yan Zhu, and Richard Zhang.
\newblock Data attribution for text-to-image models by unlearning synthesized
  images.
\newblock {\em Advances in Neural Information Processing Systems},
  37:4235--4266, 2024.

\bibitem{wu2023self}
Yiming Wu.
\newblock Self-supervised disentanglement of harmonic and rhythmic features in
  music audio signals.
\newblock {\em arXiv preprint arXiv:2309.02796}, 2023.

\bibitem{wu2021omnizart}
Yu-Te Wu, Yin-Jyun Luo, Tsung-Ping Chen, I-Chieh Wei, Jui-Yang Hsu, Yi-Chin
  Chuang, and Li~Su.
\newblock Omnizart: A general toolbox for automatic music transcription.
\newblock {\em Journal of Open Source Software}, 6(68):3391, 2021.

\end{thebibliography}
}

%%%%%%%%%%%%%%%%%%%%%%%%%%%%%%%%%%%%%%%%%%%%%%%%%%%%%%%%%%%%

\appendix
\newpage
% ============================================================
\section{The Definitions and Formulations of Methods in \ARIA}
\label{adxsec:method}
% ============================================================

\subsection{Attribution Method Formulations}
\label{app:attribution_formulations}

Each attribution method assigns a real-valued score to every training segment
$s$ given a query $q$ using per-segment loss gradients.
Let $\theta^*$ denote trained model parameters for a given stage,
$\ell(x;\theta)$ the per-segment loss, and
$g(x) = \nabla_\theta \ell(x;\theta^*) \in \mathbb{R}^d$ the loss gradient.
For each layer $l$ with weight $W_l \in \mathbb{R}^{d_l^\mathrm{out} \times d_l^\mathrm{in}}$,
the gradient has Kronecker structure:
$\mathrm{vec}(D_{W_l}) = x_l^\mathrm{in} \otimes Dx_l^\mathrm{out}$,
where $x_l^\mathrm{in}$ is the layer input and $Dx_l^\mathrm{out}$ is the output gradient.

\paragraph{\Trak.}
\Trak~\citep{park2023trak} approximates leave-one-out influence via the projected
empirical Fisher with regularization.
A dense random projection $P \in \mathbb{R}^{m \times d}$ with $m{=}4096$ gives
$\phi(x) = P g(x)$, and the score is
\begin{equation}
\label{eq:trak}
  \tau_{\Trak}(q, s)
  \;=\;
  \phi(q)^\top \bigl(\Phi\Phi^\top + \lambda I\bigr)^{-1} \phi(s),
\end{equation}
where $\Phi = [\phi(s_1),\ldots,\phi(s_M)] \in \mathbb{R}^{m \times M}$
and $\lambda$ is selected on the validation set ($\lambda{=}0.01$ in our experiments; see \Cref{app:hyperparam} for the full sweep).

\paragraph{\GradCos.}
\GradCos~\citep{charpiat2019input} omits the Hessian term and computes cosine
similarity of projected gradients:
\begin{equation}
\label{eq:gradcos}
  \tau_{\GradCos}(q, s)
  \;=\;
  \frac{\phi(q)^\top \phi(s)}{\lVert\phi(q)\rVert \lVert\phi(s)\rVert},
  \quad \phi(x) = P g(x),\; P \in \mathbb{R}^{m \times d}.
\end{equation}

\paragraph{TracIn.}
TracIn~\citep{pruthi2020estimating} accumulates projected-gradient inner products across $C$ training-time checkpoints $\theta_1,\ldots,\theta_C$:
\begin{equation}
\label{eq:tracin}
  \tau_{\mathrm{TracIn}}(q, s)
  \;=\;
  \sum_{c=1}^{C} \eta_c \, \phi_c(q)^\top \phi_c(s),
  \quad \phi_c(x) = P\, \nabla\ell(x;\theta_c),
\end{equation}
where $\eta_c$ is the learning rate at checkpoint $c$. Used only in the symbolic experiment.

\paragraph{GradDot.}
GradDot is the unnormalized single-checkpoint counterpart of \GradCos:
\begin{equation}
\label{eq:graddot}
  \tau_{\mathrm{GradDot}}(q, s)
  \;=\;
  \phi(q)^\top \phi(s),
  \quad \phi(x) = P\, \nabla\ell(x;\theta^*).
\end{equation}
Used only in the symbolic experiment through \texttt{dattri}'s \texttt{TracInAttributor} with a single checkpoint and \texttt{normalized\_grad=False}.

\paragraph{\LoGra.}
\LoGra~\citep{choe2024logra} uses a Kronecker-factored projection $P = P_i \otimes P_o$ where $P_i \in \mathbb{R}^{k_i \times d_l^\mathrm{in}}$ and $P_o \in \mathbb{R}^{k_o \times d_l^\mathrm{out}}$ are set to the leading KFAC eigenvectors of the forward and backward covariance matrices. The score is
\begin{equation}
\label{eq:logra}
  \tau_{\LoGra}(q, s)
  \;=\;
  \phi(q)^\top \bigl(P H P^\top + \delta I\bigr)^{-1} \phi(s),
  \quad \phi(x) = P g(x),
\end{equation}
where $H$ is the per-layer KFAC-approximated empirical Fisher information matrix (eFIM). The damping term $\delta = c\,\bar{\lambda}$ is scaled by the per-layer mean eigenvalue $\bar{\lambda} = \mathrm{tr}(P H P^\top)/d$, where $d$ is the projected layer dimension. The multiplier $c$ ($c{=}0.1$ in our experiments) and other hyperparameters are reported in \Cref{tab:attribution_hyperparams}. Setting $k_i \approx k_o \approx \sqrt{k}$ reduces gradient projection cost from $\mathcal{O}(dk)$ to $\mathcal{O}(\sqrt{dk})$ per segment.

\paragraph{\FactGraSS.}
\FactGraSS~\citep{hu2025grass} applies the same influence formula but replaces the
dense projection with a sparse SJLT (Sparse Johnson-Lindenstrauss Transform)
applied per Kronecker factor:
$\tilde P = \tilde P_i \otimes \tilde P_o$ where each factor is a sparse SJLT matrix.
A sparsification step first masks small gradient entries, then the sparse projection is applied:
\begin{equation}
\label{eq:factgrass}
  \tau_{\FactGraSS}(q, s)
  \;=\;
  \tilde\phi(q)^\top \bigl(\tilde\Phi\tilde\Phi^\top + \delta I\bigr)^{-1} \tilde\phi(s),
  \quad \tilde\phi(x) = \tilde P g(x).
\end{equation}
A blowup factor $b{=}4$ compensates for energy loss from gradient sparsification.
Unlike \LoGra, \FactGraSS does not use a Hessian preconditioner in the projection; regularization is applied only in the final solve.

\paragraph{Embedding baselines.}
For CLAP, CLEWS, and MERT, the score is cosine similarity of fixed audio embeddings:
\begin{equation}
\label{eq:emb}
  \tau_{\mathrm{emb}}(q, s)
  \;=\;
  \frac{e(q)^\top e(s)}{\lVert e(q)\rVert \lVert e(s)\rVert},
\end{equation}
where $e(\cdot)$ is the respective encoder's output embedding.

\subsection{Per-query normalization.}
\label{app:query-norm}
The per-query normalization in \Cref{subsec:problem-setting} is applied
only in the audio experiment, since symbolic training instances are
segment-level and require no track aggregation.
The choice of normalization follows the score distribution of each
method: \Trak and \GradCos produce bounded, near-symmetric columns
(\Trak by kernel construction; \GradCos because it is a cosine
similarity), so \textbf{z-score per test query} is used.
\LoGra and \FactGraSS are influence-function methods whose raw scores
span several orders of magnitude across queries with heavy-tailed
outliers, so \textbf{rank per test query} is used instead, as it is
outlier-resistant by construction.
TracIn and GradDot are used only in the symbolic domain experiment and
therefore require no per-query normalization.

\subsection{Feature Extraction and Similarity Definitions} 
\label{app:features}                                                                                                       
\Cref{tab:feature_params} lists the extraction tool, parameters, and output dimensionality for each feature used in the three audio channels of \Cref{sec:channels}, and \Cref{tab:feature_params_symbolic} does the same for the five symbolic channels.

\paragraph{Similarity functions.}
For each vector- or histogram-valued feature, similarity is computed using a standardized Euclidean distance~\citep{mardia1979multivariate}, where each dimension is rescaled by its standard deviation. The distance is then mapped to a similarity in $[0,1]$ via $\mathrm{sim}(a,b) = 1/(1+\|z_a-z_b\|_2)$. The standard deviations are computed once per experiment over all tracks for which features were extracted, so that the measurement scale is consistent across the dataset. They are fixed before any query is issued and shared across all queries. The audio chord-progression feature uses a normalized longest-common-subsequence score over root-motion intervals (mod 12), which is transposition-invariant.
  
\begin{table}[H]
\centering
\caption{Feature extraction parameters for the five symbolic evidence channels, with jSymbolic 2.2~\citep{mckay2018jsymbolic} from decoded MIDI. Segment duration set by dattri benchmark.}
\label{tab:feature_params_symbolic}
\vskip 0.1in
\footnotesize
\begin{tabular}{llp{6.0cm}c}
\toprule
Channel & Feature (jSymbolic ID) & Description & Dim \\
\midrule
Melody
  & Melodic-interval histogram (M-1)
    & Counts of $\pm 12$ semitone intervals
    & 25 \\
\midrule
\multirow{2}{*}{Harmony}
  & Pitch-class histogram (P-2)
    & Count per pitch class, $L_1$-normalized
    & 12 \\[2pt]
  & Vertical-interval histogram (C-2)
    & Pairwise interval classes on a 50\,Hz grid
    & 12 \\
\midrule
\multirow{3}{*}{Rhythm}
  & Note density (R-3)
    & Notes per second
    & 1 \\[2pt]
  & Mean rhythmic value (R-15)
    & Average note duration in beats
    & 1 \\[2pt]
  & Rhythmic-value histogram (R-23)
    & Counts over 32nd to double-whole values
    & 11 \\
\midrule
\multirow{3}{*}{Dynamic}
  & Average $|\Delta\mathrm{velocity}|$ (D-1)
    & Mean absolute velocity change between consecutive onsets
    & 1 \\[2pt]
  & Velocity standard deviation (D-2)
    & Standard deviation of velocity
    & 1 \\[2pt]
  & Dynamic range (D-3)
    & $\max - \min$ velocity
    & 1 \\
\midrule
\multirow{3}{*}{Texture}
  & Mean polyphony (T-2)
    & Average simultaneously-sounding notes (50\,Hz grid)
    & 1 \\[2pt]
  & Polyphony standard deviation (T-7)
    & Standard deviation of the same grid count
    & 1 \\[2pt]
  & Pitch range (P-6)
    & $\max - \min$ pitch in semitones
    & 1 \\
\bottomrule
\end{tabular}
\vskip -0.1in
\end{table}

\begin{table}[H]
\centering
\caption{Feature extraction parameters for the three evidence channels (audio at 48\,kHz mono).}
\label{tab:feature_params}
\vskip 0.1in
\footnotesize
\begin{tabular}{llp{5.2cm}c}
\toprule
Channel & Feature & Parameters & Dim \\
\midrule
\multirow{2}{*}{Rhythmic}
  & Beat interval histogram
    & BeatNet~\citep{heydari2021beatnet} offline/DBN; 16 bins $\in[0.2,\,2.0]$\,s
    & 16 \\[2pt]
  & Onset interval histogram
    & librosa~\citep{mcfee2015librosa} onset\_detect; 16 bins $\in[0.05,\,1.0]$\,s
    & 16 \\
\midrule
\multirow{3}{*}{Harmonic}
  & Chroma similarity
    & librosa chroma\_cqt; 12-bin mean
    & 12 \\[2pt]
  & Tonnetz similarity
    & librosa tonnetz~\citep{muller2015fundamentals,harte2006detecting}; from chroma
    & 6 \\[2pt]
  & Chord progression similarity
    & Omnizart~\citep{wu2021omnizart} chord\_v1; LCS on root-motion intervals (mod\,12)
    & n/a \\
\midrule
\multirow{2}{*}{Timbral}
  & MFCC similarity
    & librosa MFCC~\citep{davis1980mfcc}; $n_\mathrm{mfcc}{=}13$; $[\mu\|\sigma]$
    & 26 \\[2pt]
  & CQT similarity
    & librosa CQT; $n_\mathrm{bins}{=}84$; amplitude\_to\_dB; $[\mu\|\sigma]$
    & 168 \\
\bottomrule
\end{tabular}
\vskip -0.1in
\end{table}

% ============================================================
\section{Further Experiment Details for Reproducibility}
\label{adxsec:experiments}
% ============================================================

\subsection{Symbolic Experimental Details}

\paragraph{Model and dataset.}
We use the \texttt{dattri} benchmark to obtain a pre-trained MusicTransformer~\citep{huang2019music} on
MAESTRO~\citep{hawthorne2018maestro} together with the published LDS ground truth, and
perform no additional training. The benchmark provides $5{,}000$ training segments
and $178$ test queries per attribution method, with each segment a fixed-length
tokenized MIDI block.

\paragraph{Attribution methods.}
\Trak (10-ensemble), TracIn, \GradCos, and GradDot are run through
\texttt{dattri}'s \texttt{TRAKAttributor} and \texttt{TracInAttributor} classes
with the library defaults. Our reproduced LDS values match the benchmark's
reported numbers to $\leq 10^{-3}$. Method formulations are given in
\Cref{app:attribution_formulations}.

\subsection{Audio Experimental Details}

\paragraph{Model architecture.}
All three stages share the same transformer architecture (dim $1024$, depth $24$,
$16$ heads / $8$ KV heads, head dim $128$).
Table~\ref{tab:model_arch} summarizes the stage-specific input duration, prediction
target, output rate, number of quantizers, and conditioner.

\begin{table}[H]
\centering
\caption{Per-stage model configuration.}
\label{tab:model_arch}
\small
\setlength{\tabcolsep}{5pt}
\begin{tabular}{lccccl}
\toprule
Stage & Input & Predicts & Output Hz & \# Quant. & Conditioner \\
\midrule
Semantic & $10\,\mathrm{s}$ & MERT k-means tokens & 50 & 1024 classes & CLAP-RVQ \\
Coarse   & $4\,\mathrm{s}$  & EnCodec RVQ  & 75 & 3 & CLAP-RVQ, semantic \\
Fine     & $2\,\mathrm{s}$  & EnCodec RVQ  & 75 & 5 & CLAP-RVQ, semantic, coarse \\
\bottomrule
\end{tabular}
\end{table}

CLAP audio embeddings are quantized with a 12-codebook RVQ (codebook size $1024$)
before being passed as conditioning tokens to all stages.

\paragraph{Training setup.}
All training uses AdamW with cosine learning-rate decay and weight decay $0.01$.
Table~\ref{tab:training_setup} lists the per-stage hyperparameters.

\begin{table}[H]
\centering
\caption{Per-stage training hyperparameters.}
\label{tab:training_setup}
\small
\begin{tabular}{lrrrr}
\toprule
Stage & Steps & LR & Warmup & Eff.\ batch \\
\midrule
Semantic & 30{,}000   & $1{\times}10^{-4}$ & 1{,}000 & 40 \\
Coarse   & 100{,}000  & $2{\times}10^{-4}$ & 300     & 16 \\
Fine     & 100{,}000  & $3{\times}10^{-4}$ & 3{,}000 & 20 \\
\bottomrule
\end{tabular}
\end{table}

\paragraph{Dataset.}
We use FMA Large~\citep{defferrard2017fma}, which provides 30-second tracks. All audio is resampled to $24\,\mathrm{kHz}$.
Our implementation is based on the open-musiclm
codebase\footnote{\url{https://github.com/zhvng/open-musiclm}}, and we follow
its preprocessing pipeline without modification.
Two quality filters are applied:
(1) tracks in the Experimental genre (FMA genre ID 38) with fewer than
$1{,}000$ listens or fewer than $5$ favorites are removed, as this subset
contains a disproportionate share of noise-like content;
(2) tracks with more than $15\,\mathrm{s}$ of silence are removed.
After filtering, the dataset contains $N_\mathrm{train}{=}67{,}219$ training
tracks and $7{,}148$ held-out evaluation tracks.
Each training track is divided into non-overlapping segments of the
stage-specific audio length: $10\,\mathrm{s}$ (semantic), $4\,\mathrm{s}$
(coarse), and $2\,\mathrm{s}$ (fine), yielding
$M_\mathrm{sem}{=}200{,}798$, $M_\mathrm{crs}{=}469{,}347$, and
$M_\mathrm{fin}{=}1{,}005{,}520$ segments respectively.
Attribution is computed for each stage independently over its own segment set.

\paragraph{Embedding baselines.}
The three retrieval baselines use publicly released checkpoints.
\CLAP~\citep{elizalde2023clap} uses the LAION-CLAP
music+speech+audioset variant
(\texttt{music\_speech\_audioset\_epoch\_15\_esc\_89.98.pt}).
\CLEWS uses the SHS-trained checkpoint (\texttt{shs-clews}) released
with~\citep{serra2025clews}.
\MERT~\citep{li2023mert} uses \texttt{m-a-p/MERT-v1-95M}.
For each baseline, audio is resampled to the encoder's expected rate and the
output embedding is averaged across the temporal axis to produce a single
fixed-dimensional vector per track.

\paragraph{Attribution hyperparameters.}
Table~\ref{tab:attribution_hyperparams} summarizes the hyperparameter choices
for each attribution method. Where published defaults exist, we use them
without modification. \FactGraSS{} has no published default for the damping
term and is tuned via cross-validation in the original paper. We use
$10^{-2} \times \bar\lambda$, which lies within the searched grid.
\begin{table}[H]
\centering
\caption{Attribution method hyperparameters. Default sources: P denotes the
original paper, D the \texttt{dattri} library~\citep{deng2024dattri}.}
\label{tab:attribution_hyperparams}
\small
\begin{tabular}{lllcl}
\toprule
Method & Parameter & Default & Source & Experiment setting \\
\midrule
\Trak & Projection dim & 512 & D & 4096 \\
      & Regularization $\lambda$ & 0 & P & 0.01 \\
\GradCos & Projection dim & 512 & D & 4096 \\
\LoGra & Projection dim & 4096 ($=64^{2}$) & D & 4096 \\
       & Hessian & eFIM & P & eFIM \\
       & Damping $\delta$ & $0.1 \times \bar\lambda$ & P & $0.1 \times \bar\lambda$ \\
\FactGraSS & Projection dim & 4096 & D & 4096 \\
           & Hessian & eFIM & D & eFIM \\
           & Damping $\delta$ & $10^{-5} \times \bar\lambda$ & D & $10^{-2} \times \bar\lambda$ \\
           & Blowup factor $b$ & 4 & P/D & 4 \\
\bottomrule
\end{tabular}
\end{table}

\paragraph{Computational cost.}
\Cref{tab:compute} reports the end-to-end cost of the four attribution methods at the canonical hyperparameter settings used in our main results.

\begin{table}[H]
\centering
\caption{Compute and storage cost per attribution method on $4\!\times\!$A100-40GB, run end-to-end across the three stages in sequence at projection dimension 4096 (\Trak\ $\lambda{=}0.01$, \FactGraSS\ damping $10^{-2}$, \LoGra\ damping $0.1$). Wall-clock is elapsed time. GPU-hours bills each parallel shard separately. Peak disk is the maximum on-disk working set, including gradient cache and score matrices, with no cleanup between stages.}
\label{tab:compute}
\setlength{\tabcolsep}{8pt}
\small
\begin{tabular}{l r r r}
\toprule
Method      & Wall-clock & GPU-hours & Peak disk \\
\midrule
\Trak       & 22\,h      &  75       & 136\,GB    \\
\FactGraSS  & 14\,h      &  50       & 256\,GB    \\
\GradCos    & 29\,h      & 110       & 140\,GB    \\
\LoGra      & 19\,h      &  74       & 2.1\,TB    \\
\midrule
Total       & 84\,h      & 309       & --         \\
\bottomrule
\end{tabular}
\end{table}

\subsection{\Trak Hyperparameter Sensitivity}
\label{app:hyperparam}
\Cref{tab:hyperparam_sensitivity} reports $\bar{z}$ at $K{=}300$ over projection dimension and $\lambda$ for all nine stage-channel cells. At $\lambda{=}0$, fine harmonic is strongly negative at both projection dimensions ($-5.52$ at 2048, $-5.70$ at 4096). Any $\lambda \in [10^{-2}, 10^{2}]$ recovers it to between $+1.25$ and $+1.31$.
% Full hyperparameter sensitivity: all 9 stage × channel combinations.
% Values: mean z at K=300, Ntest=7,148.
% † No regularization (λ=0, pipeline default).  ★ Selected configuration.
\begin{table}[h]
\footnotesize
\centering
\caption{\Trak hyperparameter sensitivity. $\bar{z}$ at $K{=}300$ across projection dim and $\lambda$. Positive values in \textbf{bold}, selected configuration marked $(\star)$.}
\label{tab:hyperparam_sensitivity}
\vskip 0.1in
\resizebox{\linewidth}{!}{%
\begin{tabular}{ccrrrrrrrrr}
\toprule
& &
  \multicolumn{3}{c}{Semantic} &
  \multicolumn{3}{c}{Coarse} &
  \multicolumn{3}{c}{Fine} \\
\cmidrule(lr){3-5}\cmidrule(lr){6-8}\cmidrule(lr){9-11}
Proj. & $\lambda$ &
  Rhy. & Har. & Tim. &
  Rhy. & Har. & Tim. &
  Rhy. & Har. & Tim. \\
\midrule
\multirow{4}{*}{2048}
  & $0$\rlap{$^\dagger$}
    & \textbf{+1.43} & $-3.30$ & $-0.55$
    & $-1.47$ & $-2.78$ & $-3.67$
    & $-2.11$ & $-5.52$ & $-6.78$ \\
  & $10^{-2}$
    & \textbf{+1.54} & $-3.66$ & $-0.91$
    & $-2.54$ & $-3.34$ & $-7.86$
    & $-0.05$ & \textbf{+1.31} & $-0.53$ \\
  & $10^{-1}$
    & \textbf{+1.42} & $-3.16$ & $-0.81$
    & $-3.88$ & $-3.12$ & $-9.98$
    & $-0.03$ & \textbf{+1.31} & $-0.52$ \\
  & $1.0$
    & \textbf{+1.39} & $-3.03$ & $-0.84$
    & $-3.89$ & $-2.83$ & $-9.98$
    & $-0.01$ & \textbf{+1.30} & $-0.51$ \\
\midrule
\multirow{6}{*}{4096}
  & $0$\rlap{$^\dagger$}
    & \textbf{+1.40} & $-2.95$ & $-0.51$
    & $-3.87$ & $-5.17$ & $-9.07$
    & $-0.99$ & $-5.70$ & $-6.15$ \\
  & $10^{-2}$\rlap{$^{\star}$}
    & \textbf{+1.51} & $-3.63$ & $-0.86$
    & $-2.78$ & $-5.89$ & $-7.58$
    & \textbf{+0.45} & \textbf{+1.25} & \textbf{+0.73} \\
  & $10^{-1}$
    & \textbf{+1.43} & $-3.12$ & $-0.56$
    & $-3.74$ & $-5.25$ & $-8.69$
    & \textbf{+0.47} & \textbf{+1.26} & \textbf{+0.72} \\
  & $1.0$
    & \textbf{+1.40} & $-2.97$ & $-0.51$
    & $-3.86$ & $-5.18$ & $-9.03$
    & \textbf{+0.48} & \textbf{+1.28} & \textbf{+0.70} \\
  & $30$
    & \textbf{+1.40} & $-2.95$ & $-0.51$
    & $-3.87$ & $-5.17$ & $-9.07$
    & \textbf{+0.46} & \textbf{+1.27} & \textbf{+0.72} \\
  & $100$
    & \textbf{+1.40} & $-2.95$ & $-0.51$
    & $-3.87$ & $-5.17$ & $-9.08$
    & \textbf{+0.48} & \textbf{+1.30} & \textbf{+0.72} \\
\bottomrule
\end{tabular}%
}
\vskip 0.05in
\begin{minipage}{\linewidth}
  \footnotesize
  $^\dagger$~No regularization ($\lambda{=}0$, pipeline default).
  $^\star$~Selected configuration used in all main-paper experiments.
\end{minipage}
\vskip -0.1in
\end{table}

\subsection{Full $K$-Sweep results}
\label{app:full_result}
\paragraph{Symbolic Domain}
All four attribution methods on MusicTransformer have $r_1{<}0.15$ and $r_{2:5}{<}0.25$ (\Cref{tab:dattri_combined}), so the residual $\bar z_c$ stays close to the original on every channel at every $K$. The channel rankings at $K{=}300$ hold across the full sweep. TracIn's harmony remains the strongest positive signal, growing from $+1.09$ at $K{=}20$ to $+2.43$ at $K{=}500$, while GradDot's dynamics remains the strongest negative, moving from $-0.10$ to $-0.78$ over the same range.

\begin{figure}[H]
  \centering
  \begin{subfigure}[t]{0.32\textwidth}\includegraphics[width=\linewidth]{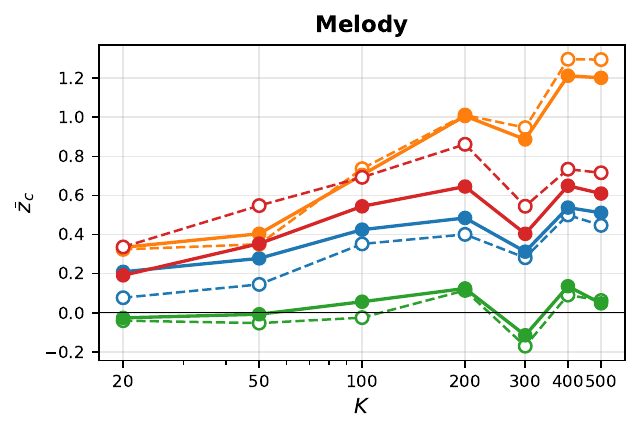}\caption{Melody}\end{subfigure}\hfill
  \begin{subfigure}[t]{0.32\textwidth}\includegraphics[width=\linewidth]{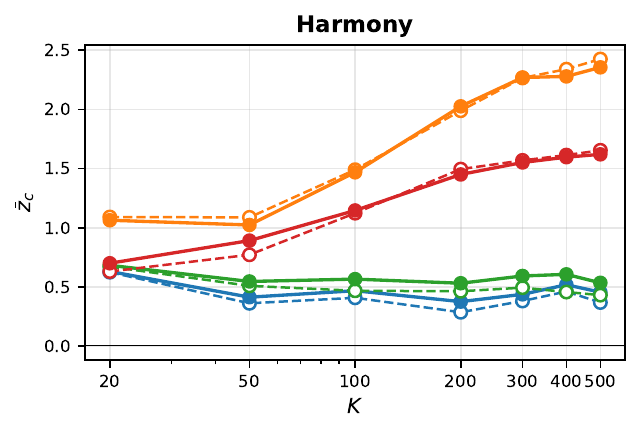}\caption{Harmony}\end{subfigure}\hfill
  \begin{subfigure}[t]{0.32\textwidth}\includegraphics[width=\linewidth]{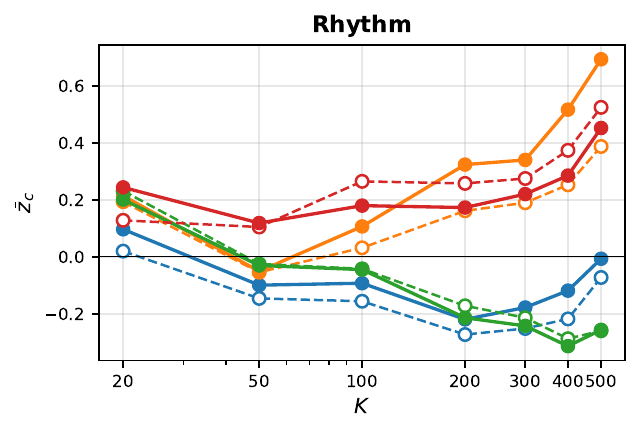}\caption{Rhythm}\end{subfigure}\\[0.4em]
  \begin{subfigure}[t]{0.32\textwidth}\includegraphics[width=\linewidth]{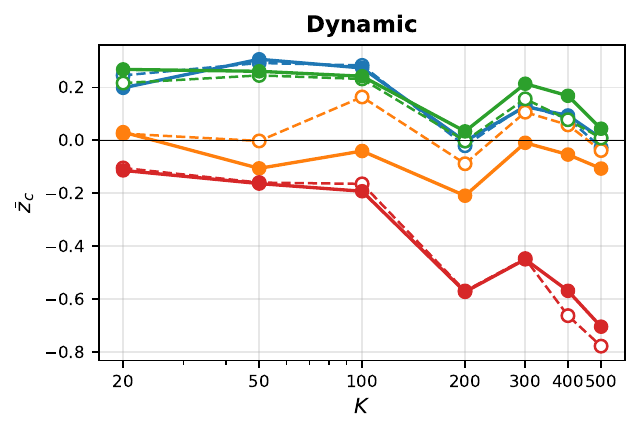}\caption{Dynamic}\end{subfigure}\hfill
  \begin{subfigure}[t]{0.32\textwidth}\includegraphics[width=\linewidth]{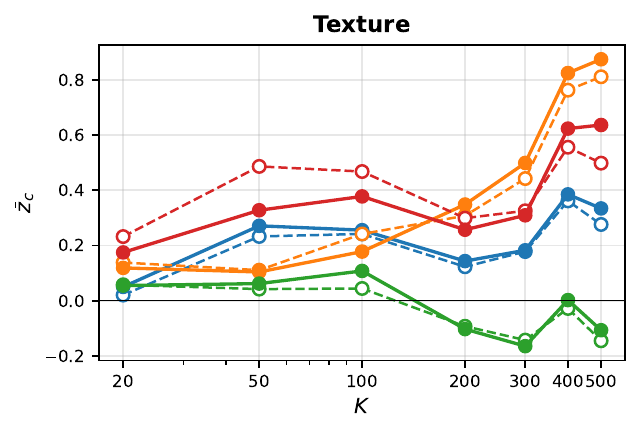}\caption{Texture}\end{subfigure}\\[0.3em]
  \includegraphics[width=0.85\textwidth]{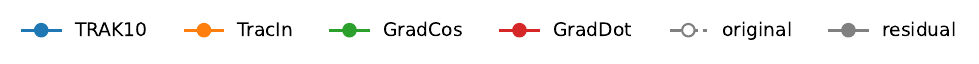}
  \caption{Within-group homogeneity $\bar z_c$ across $K$ on MusicTransformer + MAESTRO, one panel per jSymbolic channel with the four attribution methods overlaid. Dashed curves use the original $S^{\mathrm{seg}}$, solid curves use the rank-1 residual.}
  \label{fig:dattri_ksweep}
\end{figure}

\paragraph{Audio Domain}
The two figures below extend the body's \Cref{fig:ksweep_main} with the eight attribution stage-method settings not shown there and the three embedding-based retrieval baselines. Numerical $K{=}300$ values are tabulated in \Cref{tab:residual_homogeneity}.

\begin{figure}[h]
  \centering
  \begin{subfigure}[t]{0.245\textwidth}\includegraphics[width=\linewidth]{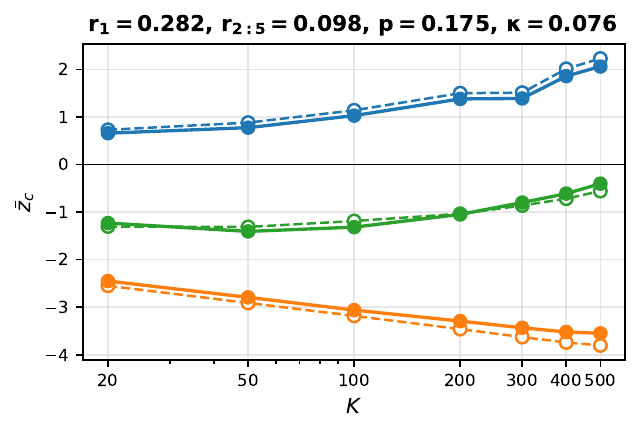}\caption{\Trak semantic}\end{subfigure}\hfill
  \begin{subfigure}[t]{0.245\textwidth}\includegraphics[width=\linewidth]{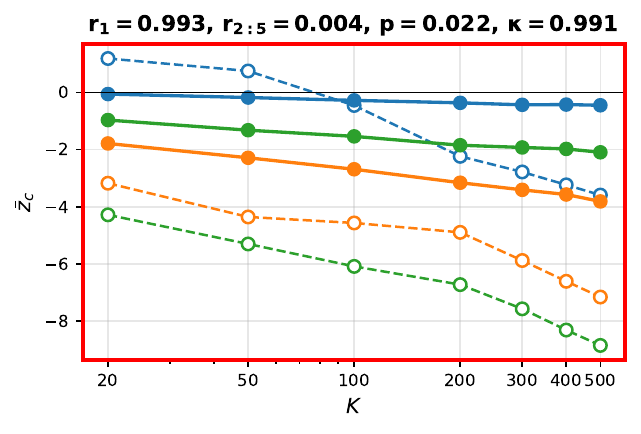}\caption{\Trak coarse}\end{subfigure}\hfill
  \begin{subfigure}[t]{0.245\textwidth}\includegraphics[width=\linewidth]{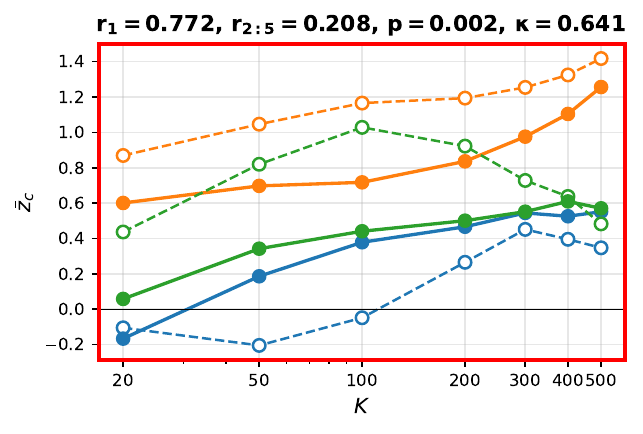}\caption{\Trak fine}\end{subfigure}\hfill
  \begin{subfigure}[t]{0.245\textwidth}\includegraphics[width=\linewidth]{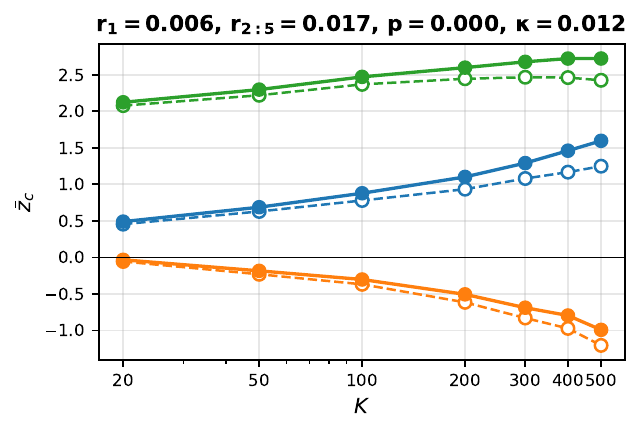}\caption{\FactGraSS semantic}\end{subfigure}\\[0.4em]
  \begin{subfigure}[t]{0.245\textwidth}\includegraphics[width=\linewidth]{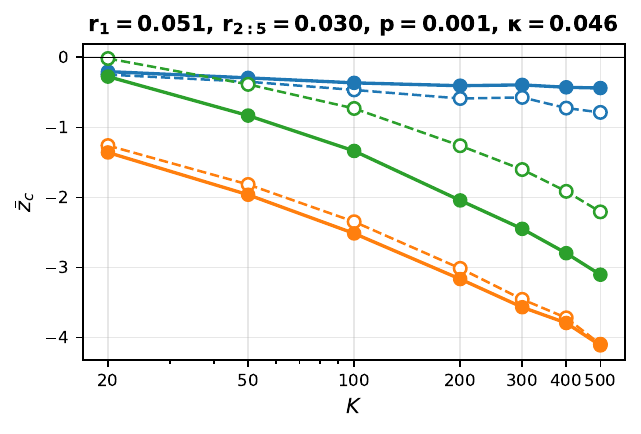}\caption{\FactGraSS coarse}\end{subfigure}\hfill
  \begin{subfigure}[t]{0.245\textwidth}\includegraphics[width=\linewidth]{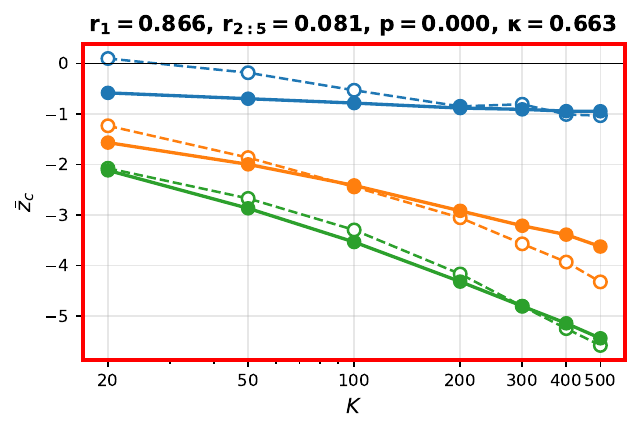}\caption{\FactGraSS fine}\end{subfigure}\hfill
  \begin{subfigure}[t]{0.245\textwidth}\includegraphics[width=\linewidth]{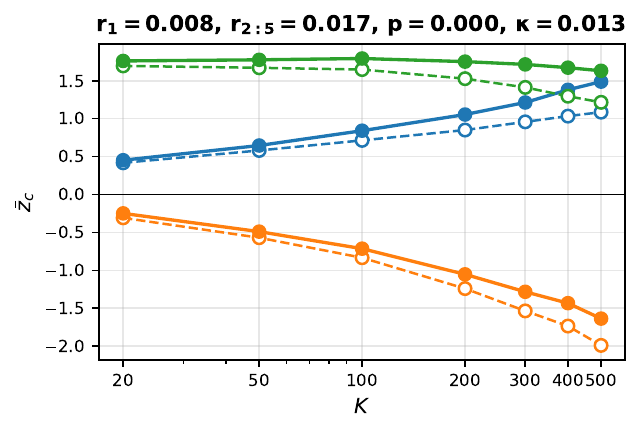}\caption{\LoGra semantic}\end{subfigure}\hfill
  \begin{subfigure}[t]{0.245\textwidth}\includegraphics[width=\linewidth]{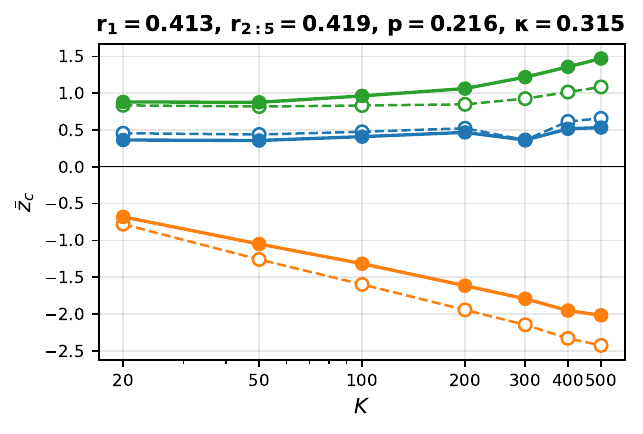}\caption{\GradCos semantic}\end{subfigure}\\[0.3em]
  \includegraphics[width=0.85\textwidth]{figs/ksweep_cells/legend.pdf}
  \caption{Homogeneity $\bar z_c$ across $K$ for the eight attribution stage-method settings not shown in \Cref{fig:ksweep_main}, with identical curve conventions.}
  \label{fig:ksweep_appendix_attr}
\end{figure}

\begin{figure}[h]
  \centering
  \begin{subfigure}[t]{0.30\textwidth}\includegraphics[width=\linewidth]{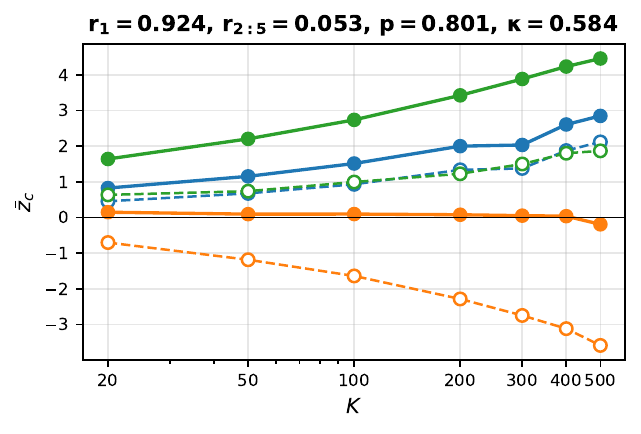}\caption{\CLAP}\end{subfigure}\hfill
  \begin{subfigure}[t]{0.30\textwidth}\includegraphics[width=\linewidth]{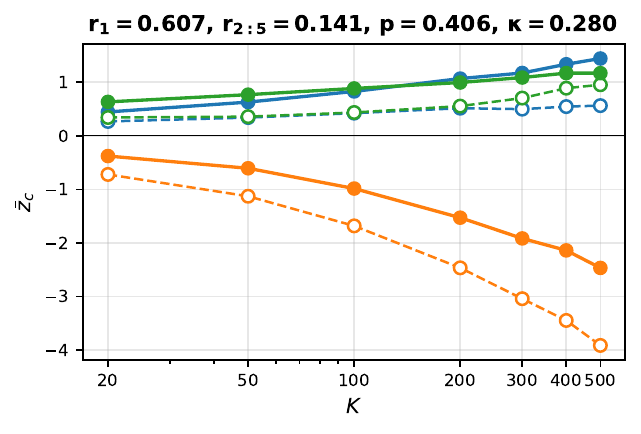}\caption{\CLEWS}\end{subfigure}\hfill
  \begin{subfigure}[t]{0.30\textwidth}\includegraphics[width=\linewidth]{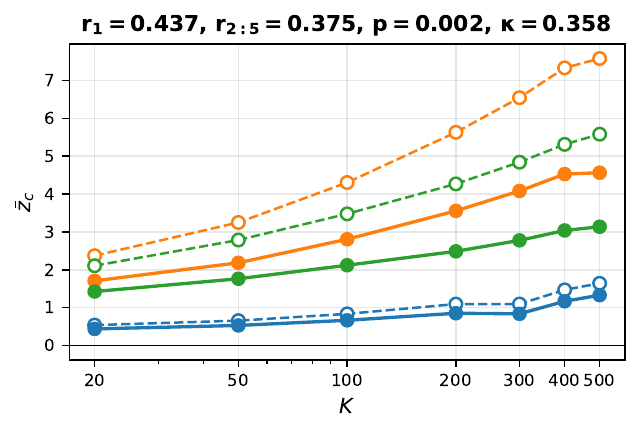}\caption{\MERT}\end{subfigure}\\[0.3em]
  \includegraphics[width=0.85\textwidth]{figs/ksweep_cells/legend.pdf}
  \caption{Homogeneity $\bar z_c$ across $K$ for the three embedding-based retrieval baselines. Curve conventions as in \Cref{fig:ksweep_main}.}
  \label{fig:ksweep_retrieval}
\end{figure}

\newpage
\subsection{Residual Homogeneity at $K{=}300$}

\Cref{tab:residual_homogeneity} reports the per-channel residual $\bar z^{\mathrm{res}}_c$, Pos, and Sig at $K{=}300$ for all 15 settings, supplementing the K-sweep curves above with exact values.

\begin{table}[h]
\centering
\caption{%
  Residual within-group musical homogeneity
  ($\bar{z}^{\mathrm{res}}_c$, Pos, Sig) after subtracting
  $\sigma_1 u_1 v_1^\top$ from $S^{\mathrm{seg}}$, for all 15 settings
  ($K{=}300$, $N_\text{test}{=}7{,}148$).
  Stage: S = semantic, C = coarse, F = fine.
  Reliability diagnostics ($r_1$, $r_{2:5}$, $p$, $\kappa$) are
  computed on the original $S^{\mathrm{seg}}$ and repeated here for
  comparison with \Cref{tab:attribution_coherence}.
  Bold marks positive $\bar{z}^{\mathrm{res}}$ and the best (lowest) reliability value within each method group.
}
\label{tab:residual_homogeneity}
\vskip 0.1in
\resizebox{\linewidth}{!}{%
\begin{tabular}{llcccc rrr rrr rrr}
\toprule
& & & & & & \multicolumn{3}{c}{Rhythmic} & \multicolumn{3}{c}{Harmonic} & \multicolumn{3}{c}{Timbral} \\
\cmidrule(lr){7-9}\cmidrule(lr){10-12}\cmidrule(lr){13-15}
Method & Stage & $r_1$ & $r_{2:5}$ & $p$ & $\kappa$
  & $\bar{z}^{\mathrm{res}}$ & Pos$\uparrow$ & Sig$\uparrow$
  & $\bar{z}^{\mathrm{res}}$ & Pos$\uparrow$ & Sig$\uparrow$
  & $\bar{z}^{\mathrm{res}}$ & Pos$\uparrow$ & Sig$\uparrow$ \\
\midrule
\multirow{3}{*}{\Trak}
  & S & \textbf{0.282} & 0.098 & 0.175 & \textbf{0.076}
    & \textbf{+1.39} & 86.9\% & 31.7\%
    & $-$3.43        &  1.9\% &  0.0\%
    & $-$0.80        & 29.6\% &  3.3\% \\
  & C & 0.993 & 0.004 & 0.022 & 0.991
    & $-$0.43        & 42.5\% & 14.4\%
    & $-$3.41        &  2.2\% &  0.1\%
    & $-$1.93        & 31.4\% & 23.9\% \\
  & F & 0.772 & 0.208 & \textbf{0.002} & 0.641
    & \textbf{+0.54} & 74.5\% &  5.1\%
    & \textbf{+0.98} & 87.5\% & 12.6\%
    & \textbf{+0.55} & 74.6\% &  4.7\% \\
\midrule
\multirow{3}{*}{\FactGraSS}
  & S & \textbf{0.006} & 0.018 & \textbf{0.000} & \textbf{0.012}
    & \textbf{+1.29} & 63.4\% & 42.6\%
    & $-$0.69        & 40.1\% & 23.0\%
    & \textbf{+2.68} & 65.9\% & 53.3\% \\
  & C & 0.051 & 0.030 & 0.001 & 0.046
    & $-$0.39        & 38.8\% & 13.1\%
    & $-$3.57        &  9.0\% &  2.7\%
    & $-$2.45        & 25.4\% & 14.0\% \\
  & F & 0.866 & 0.081 & \textbf{0.000} & 0.663
    & $-$0.91        & 18.2\% &  0.2\%
    & $-$3.21        &  0.3\% &  0.0\%
    & $-$4.80        &  0.0\% &  0.0\% \\
\midrule
\multirow{3}{*}{\GradCos}
  & S & \textbf{0.413} & 0.419 & 0.216 & \textbf{0.374}
    & \textbf{+0.36} & 55.3\% & 20.4\%
    & $-$1.79        & 25.0\% &  7.2\%
    & \textbf{+1.22} & 70.8\% & 37.3\% \\
  & C & 1.000 & 0.000 & 1.000 & 0.997
    & \textbf{+1.34} & 57.1\% & 45.4\%
    & $-$3.84        & 22.2\% &  7.2\%
    & $-$7.57        &  1.1\% &  0.0\% \\
  & F & 0.793 & 0.200 & \textbf{0.015} & 0.736
    & \textbf{+0.72} & 72.1\% & 14.5\%
    & $-$5.40        &  2.3\% &  0.2\%
    & $-$4.32        &  4.3\% &  0.4\% \\
\midrule
\multirow{3}{*}{\LoGra}
  & S & \textbf{0.008} & 0.017 & \textbf{0.000} & \textbf{0.013}
    & \textbf{+1.21} & 65.0\% & 40.1\%
    & $-$1.28        & 33.1\% & 16.1\%
    & \textbf{+1.71} & 62.0\% & 46.9\% \\
  & C & 0.057 & 0.038 & 0.006 & 0.049
    & $-$0.13        & 45.7\% & 24.3\%
    & $-$2.43        & 27.4\% & 16.0\%
    & \textbf{+1.54} & 54.0\% & 41.7\% \\
  & F & 0.223 & 0.065 & \textbf{0.000} & 0.272
    & $-$0.77        & 22.9\% &  0.5\%
    & $-$3.39        &  0.2\% &  0.0\%
    & $-$4.68        &  0.0\% &  0.0\% \\
\midrule
CLAP  & --- & 0.924 & 0.054 & 0.801 & 0.584
    & \textbf{+2.03} & 57.5\% & 45.0\%
    & \textbf{+0.05} & 48.3\% & 40.2\%
    & \textbf{+3.89} & 61.6\% & 54.0\% \\
CLEWS & --- & 0.607 & 0.141 & 0.406 & \textbf{0.280}
    & \textbf{+1.17} & 58.9\% & 36.3\%
    & $-$1.92        & 43.4\% & 32.4\%
    & \textbf{+1.09} & 57.5\% & 43.7\% \\
MERT  & --- & \textbf{0.437} & 0.375 & \textbf{0.002} & 0.358
    & \textbf{+0.84} & 59.0\% & 34.2\%
    & \textbf{+4.08} & 76.4\% & 63.9\%
    & \textbf{+2.78} & 76.8\% & 64.6\% \\
\bottomrule
\end{tabular}%
}
\vskip -0.1in
\end{table}

\subsection{Genre Confound Analysis}
\label{app:genre_confound}

Since random reference groups are sampled uniformly from $D$, an attributed group concentrated in a single genre may inherit that genre's intrinsic within-group similarity rather than reflect query-specific attribution. We probe this confound for the canonical \Trak (proj $4096$, $\lambda{=}0.01$) fine-stage results at $K{=}300$ via a within-genre breakdown of the fine harmonic z-score.

\Cref{tab:genre_breakdown} stratifies the fine harmonic z-score by the FMA top-level genre of the test query, restricted to the $3{,}605$ genre-labelled queries. Mean\,z is the per-genre average z-score against $B{=}200$ random reference groups, Pos is the fraction of queries with z${>}0$, and Sig is the fraction with z${>}1.96$. Mean\,z is positive in all 13 genres, with Pos $\geq 87\%$ in every genre and a narrow range across genres (SD $0.11$). A confound in which the positive signal arose from a single dominant genre would predict large between-genre variance, but the observed across-genre standard deviation of $0.11$ is inconsistent with that explanation.

\begin{table}[h]
\footnotesize
\centering
\caption{%
  Fine-stage harmonic z-score stratified by FMA top-level genre at $K{=}300$
  ($n{=}3{,}605$ genre-labelled queries).
}
\label{tab:genre_breakdown}
\vskip 0.1in
\small
\begin{tabular}{lrrrr}
\toprule
Genre & $n$ & Mean\,z & Pos$\uparrow$ & Sig$\uparrow$ \\
\midrule
Classical          &   87 & \textbf{+1.451} & 98.9\% & 26.4\% \\
Folk               &  293 & \textbf{+1.357} & 96.9\% & 20.8\% \\
International      &  123 & \textbf{+1.343} & 97.6\% & 13.0\% \\
Instrumental       &  287 & \textbf{+1.320} & 95.8\% & 20.2\% \\
Old-Time / Historic &  54 & \textbf{+1.290} & 92.6\% & 20.4\% \\
Pop                &  194 & \textbf{+1.248} & 94.3\% & 18.6\% \\
Rock               & 1325 & \textbf{+1.241} & 94.3\% & 16.6\% \\
Hip-Hop            &  312 & \textbf{+1.222} & 94.6\% & 16.0\% \\
Electronic         &  804 & \textbf{+1.166} & 89.9\% & 16.4\% \\
Spoken             &   28 & \textbf{+1.151} & 96.4\% & 10.7\% \\
Experimental       &   20 & \textbf{+1.140} & 90.0\% & 15.0\% \\
Jazz               &   46 & \textbf{+1.068} & 91.3\% &  8.7\% \\
Soul-R\&B          &   32 & \textbf{+1.057} & 87.5\% & 18.8\% \\
\bottomrule
\end{tabular}
\vskip -0.1in
\end{table}

%%%%%%%%%%%%%%%%%%%%%%%%%%%%%%%%%%%%%%%%%%%%%%%%%%%%%%%%%%%%

\end{document}